\documentclass[traditabstract]{aa} 

\usepackage[fleqn]{amsmath}
\usepackage{amssymb}
\usepackage{textcomp}
\usepackage{graphicx}
\usepackage{txfonts}
\usepackage{natbib}
\usepackage{url}
\usepackage{multirow}
\usepackage{subfigure}
\usepackage[usenames,dvipsnames]{xcolor}
\usepackage{bm}

\newcommand{\ab}[1]{x_{\rm #1}}
\newcommand{\op}[1]{{\rm OPR(#1)}}

\newcommand{\reacteq}[4]{{\rm #1} + {\rm #2} \rightleftharpoons {\rm #3} + {\rm #4}} 

\newcommand{\hdo}{{\rm HDO/H_{2}O}}
\newcommand{\ddo}{{\rm D_{2}O/HDO}}

\newcommand{\dhr}[1]{{\rm [D/H]}_{\rm #1}}

\newcommand{\req}[3]{#1_{#2}^{(#3)}}
\newcommand{\sumreq}[2]{#1^{(#2)}}

\begin{document}
\title{Water delivery from cores to disks: deuteration as a probe of the prestellar inheritance of H$_2$O}
\titlerunning{Water delivery from cores to disks}
\authorrunning{Furuya et al.}

\author{K. Furuya\inst{\ref{inst1},\ref{inst2}} 
\and M. N. Drozdovskaya\inst{\ref{inst1}} 
\and R. Visser\inst{\ref{inst3}}
\and E. F. van Dishoeck\inst{\ref{inst1},\ref{inst4}} 
\and C. Walsh\inst{\ref{inst1},\ref{inst5}}
\and D. Harsono\inst{\ref{inst6}}
\and U. Hincelin\inst{\ref{inst7}}
\and V. Taquet\inst{\ref{inst1}}
}

\institute{Leiden Observatory, Leiden University, P.O. Box 9513, 2300 RA, The Netherlands\\
\email{furuya@ccs.tsukuba.ac.jp}\label{inst1} 
\and Center for Computational Sciences, University of Tsukuba, 1-1-1 Tennoudai, Tsukuba 305-8577, Japan\label{inst2}
\and European Southern Observatory, Karl-Schwarzschild-Str. 2, 85748 Garching, Germany\label{inst3}
\and Max-Planck-Institut f\"ur Extraterrestrische Physik, Giessenbachstrasse, 85741 Garching, Germany\label{inst4}
\and School of Physics and Astronomy, University of Leeds, Leeds, LS2 9JT, UK\label{inst5}
\and Heidelberg University, Center for Astronomy, Institute of Theoretical Astrophysics, Albert-Ueberle-Stra{\ss}e 2, 69120 Heidelberg, Germany\label{inst6}
\and Department of Chemistry, University of Virginia, Charlottesville, VA 22904, USA\label{inst7}
}


 
\abstract
{We investigate the delivery of regular and deuterated forms of water from prestellar cores to circumstellar disks.
We adopt a semi-analytical axisymmetric two-dimensional collapsing core model with post-processing gas-ice astrochemical simulations, 
in which a layered ice structure is considered.
The physical and chemical evolutions are followed until the end of the main accretion phase.
When mass averaged over the whole disk, a forming disk has a similar H$_2$O abundance and $\hdo$ abundance ratio as their precollapse values (within a factor of 2), regardless of time in our models. 
Consistent with previous studies, our models suggest that interstellar water ice is delivered to forming disks without significant alteration.
On the other hand, the local vertically averaged H$_2$O ice abundance and $\hdo$ ice ratio can differ more, by up to a factor of several, depending on time and distance from a central star.
Key parameters for the local variations are the fluence of stellar UV photons en route into the disk and the ice layered structure, 
the latter of which is mostly established in the prestellar stages.
We also find that even if interstellar water ice is destroyed by stellar UV and (partly) reformed prior to disk entry, 
the $\hdo$ ratio in reformed water ice is similar to the original value.
This finding indicates that some caution is needed in discussions on the prestellar inheritance of H$_2$O based on comparisons between 
the observationally derived $\hdo$ ratio in clouds/cores and that in disks/comets.
Alternatively, we propose that the ratio of $\ddo$ to $\hdo$ better probes the prestellar inheritance of H$_2$O.
It is also found that icy organics are more enriched in deuterium than water ice in forming disks.
The differential deuterium fractionation in water and organics is inherited from the prestellar stages.
}


\maketitle

\section{Introduction}
\label{sec:intro}
A molecular cloud core is the formation site of stars.
The gravitational collapse of cores leads to the birth of protostars, which are accompanied 
by surrounding disk-like structures, termed circumstellar disks.
When star-disk systems are born, they are deeply embedded in the surrounding core,
and evolve via the accretion of remnant material.
In disks, grain growth and/or gravitational instability eventually lead to the formation of planets, 
although the timing of planet formation in the sequence of the physical evolution remains an open question \citep[e.g.,][]{shu87,testi14,alma15,yen16}.

It is known that water ice is already abundant in molecular clouds \citep{whittet93}.
One of the major goals in the field of astrochemistry is to reveal the water trail from its formation in molecular clouds and cores
to the delivery to planetary systems \citep[][for a recent review]{vandishoeck14}.
The level of deuterium fractionation in molecules (measured via the $\hdo$ and D$_2$O/H$_2$O abundance ratios for water) depends on their formation environments.
This characteristic can allow us to gain insights into the water trail by comparing the deuterium fractionation in objects at different evolutionary stages.
Recent interferometric observations have quantified the gaseous  $\hdo$ ratio in the inner hot regions ($>$100 K) 
around deeply embedded low-mass protostars, where water ice has sublimated \citep[e.g.,][]{persson14}. 
The inferred $\hdo$ ratio is $\sim$10$^{-3}$, which is similar to that in some comets in our solar system \citep[e.g.,][]{villanueva09,altwegg15}.
This similarity may imply that some cometary water originated from the embedded protostellar or earlier prestellar phases.
On the other hand, given the variation in the $\hdo$ ratio among pristine solar system materials \citep[i.e., comets and meteorites; e.g.,][]{mumma11,alexander12}, 
the situation may be more complicated.

There have been numerous efforts to understand the water deuteration in star- and planet-forming regions using numerical simulations \citep[e.g.,][]{tielens83,aikawa99}.
In this work, we focus on the delivery of regular and deuterated forms of water from prestellar cores to circumstellar disks.
The disk formation stage is the connecting point of the interstellar chemistry and disk chemistry and thus the key to understanding the connection between both phases.
Here we outline the water ice formation and deuteration in molecular clouds/cores and describe the aim of this paper. 
The potential importance of (chemical) processing in disks themselves is discussed in Section \ref{sec:discussion}.

Astrochemical models for clouds and cores in the local interstellar medium (ISM) have shown that deuteration of water ice can occur (too) efficiently owing to the low gas temperature ($\sim$10 K) and 
the availability of cosmic rays as a source of ionization. 
Both are necessary to drive deuterium fractionation through isotopic exchange reactions in the gas phase, such as $\reacteq{H_3^+}{HD}{H_2D^+}{H_2}$ \citep{watson76}.
The elemental abundance of deuterium with respect to hydrogen is $\dhr{elem}=1.5\times10^{-5}$ in the local ISM \citep[][]{linsky03}.
The primary reservoir of deuterium is HD in the dense ISM. 
At low temperatures, the isotopic exchange reactions favor deuterium transfer from HD to e.g., H$_3^+$ over the endothermic reverse direction. 
The endothermicity depends on the nuclear spin states of the reactants and the products \citep[e.g.,][]{hugo09}.
The enrichment of deuterium in e.g., H$_3^+$ is distributed to both gaseous and icy molecules through sequential chemical reactions \citep{tielens83,millar89}.
Indeed pseudo-time dependent models for prestellar cores, in which the physical conditions are constant with time, 
usually predict an $\hdo$ ice ratio of the order of $\sim$10$^{-2}$ or even larger, 
which is greater than the upper limit on the $\hdo$ ice ratio inferred from infrared observations of ices in outer cold protostellar envelopes \citep[$<(2-10)\times 10^{-3}$,][]{dartois03,parise03}.

A promising solution to this contradiction is the large gradient of deuterium fractionation during water ice formation,
originally proposed by \citet{dartois03} and recently reproposed by \citet{furuya15,furuya16b} based on gas-ice chemical simulations 
during the formation and evolution of molecular clouds.
The basic idea is that 
(i) the majority of volatile oxygen is locked up in water ice and other O-bearing molecules in molecular clouds without significant deuterium fractionation and 
(ii) at later times, in prestellar cores, water ice formation continues with reduced efficiency but with enhanced deuterium fractionation.
The enhanced deuterium fractionation in the latter stage can be triggered by a drop in the ortho-to-para nuclear spin ratio of H$_2$, CO freeze-out, 
and the attenuation of the interstellar UV field \citep[e.g.,][]{roberts02,flower06}.
In this case, a small fraction of water ice (i.e., the outer layers of ice mantles) is formed with high deuteration ratios of $\gtrsim$10$^{-2}$, 
but the $\hdo$ ratio in the entire ice mantle is not very high ($\ll$10$^{-2}$). 
This scenario is also consistent with the higher $\ddo$ ratio compared to the $\hdo$ ratio\footnote{
The production rate of X$_2$O ice is denoted as $R_{\rm X_2O}$, where X is H or D.
$R_{\rm HDO} \propto f_{\rm D}R_{\rm H_2O}$ and $R_{\rm D_2O} \propto f_{\rm D}^2 R_{\rm H_2O}$, 
where $f_{\rm D}$ is the atomic D/H ratio.
Then, the total amount of D$_2$O is sensitive to the water ice formation in the later stage when deuterium fractionation is efficient.
This makes D$_2$O a good probe of the water ice formation in the later stage.} 
measured in the inner hot regions around a Class 0 protostar, where water ice has sublimated \citep{coutens14,furuya16b}.
It also explains the higher levels of deuterium fractionation in formaldehyde and methanol than that in water in the inner hot regions around protostars \citep[e.g.,][]{parise06};
formaldehyde and methanol form via sequential hydrogenation of CO ice \citep[e,g.,][]{watanabe02} primarily in prestellar cores, 
where deuterium fractionation is enhanced \citep{cazaux11,taquet12,furuya16b}.
In summary, the chemical and isotopic compositions of the ISM ice are characterized not only by the bulk ice composition, 
but also by the differentiation within the ice mantle, reflecting the physical and chemical evolution during the prestellar stages \citep[e,g.,][]{rodgers08,boogert15}.

\citet{visser09b,visser11} studied the chemical evolution from prestellar cores to circumstellar disks with an axisymmetric two-dimensional (2D) collapse model.
They suggested that disks contain significant amounts of interstellar H$_2$O ice which has been delivered from cores without experiencing significant UV and thermal processing.
In order to test this scenario observationally, we need a chemical tracer which probes the processing of water ice, i.e., deuteration.
One may think that if the claim of Visser et al. were true, it would hold for HDO ice as well and the prestellar $\hdo$ ice ratio would be preserved.
This assumption would be correct, if the ISM ices were well-mixed in terms of chemical composition.
However, given the layered ice structure in the ISM, the situation for HDO can be different from that for H$_2$O, 
because the upper ice mantle layers, where the significant fraction of HDO is present, interact with the gas phase more easily than the lower ice mantle layers.

The aim of the present work is to investigate the delivery of layered ice from prestellar cores to forming disks with a special focus on regular and deuterated water.
We revisit the collapse model of Visser et al. with a multiphase gas-ice chemical model, in which the ice layered structure and deuteration are considered.
In terms of chemistry, the protostar/disk formation may be divided into three stages:
(i) processing of prestellar materials (gas and ice) by stellar heating and UV photons \citep{vandishoeck98}, 
(ii) accretion shock heating upon disk entry \citep[][]{lunine91,sakai14}, and
(iii) chemical evolution and redistribution of angular momentum (i.e., materials) in the forming disk.
The main focus of this work is stage (i), although our model follows the physical and chemical evolution in the disk as well.
The accretion shock heating is not taken into account in our model; 
for the dust grain size assumed in the present work (radius of 0.1 $\mu$m), the stellar heating of dust grains dominates over the shock heating \citep{visser09b}.
Preliminary results of this work are in part presented in \citet{furuya16a}, which will appear in the proceedings of the 6th Zermatt ISM Symposium.

This paper is organized as follows.
We describe our physical and chemical models in Section \ref{sec:phys} and \ref{sec:chem}, respectively.
In Section \ref{sec:result} we present the structure of water ice abundances and the deuteration ratios 
in an infalling envelope and in a forming disk in our fiducial model.
Parameter dependences are discussed in Section \ref{sec:parameter}.
The implication of our results to arguments on the origin of disk/cometary water is addressed in Section \ref{sec:discussion}.
Differences between the level of deuteration in water and that in organics are also discussed.
Our findings are summarized in Section \ref{sec:summary}. 

\section{Physical model}
\label{sec:phys}

\begin{figure*}
\sidecaption
\resizebox{\hsize}{9cm}{\includegraphics{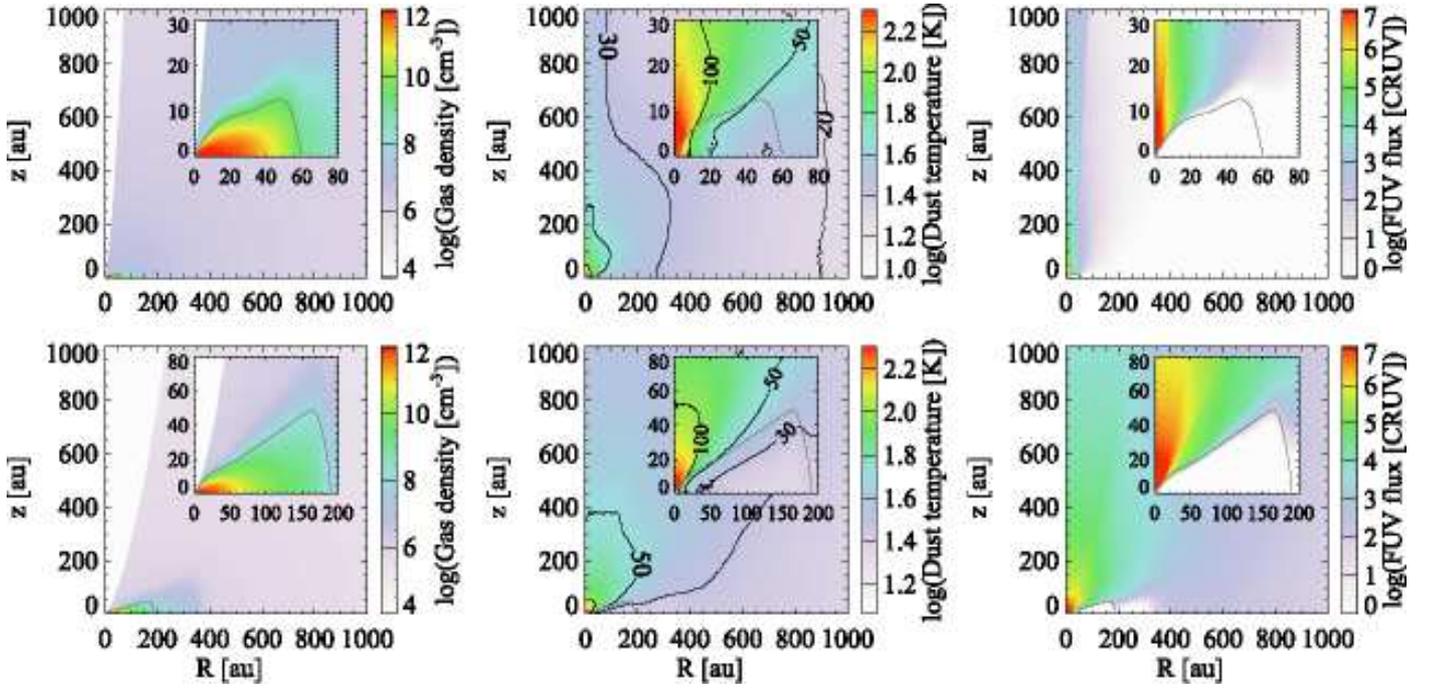}}
\caption{Spatial distributions of the gas density (left), dust temperature (middle), and stellar FUV radiation field 
normalized by cosmic ray-induced radiation field ($10^4$ photons cm$^{-2}$ s$^{-1}$; right) 
at $t = 0.5 t_{\rm acc}$ (top) and $0.9t_{\rm acc}$ (bottom) in our fiducial model.
In the middle panel, contours are drawn at 100 K, 50 K, 30 K, and 20 K.
Insets zoom in on the disk. The disk surface is shown by the gray solid line.}
\label{fig:phys}
\end{figure*}

We simulate the physical evolution from the collapse of rotating prestellar cores to the formation of circumstellar disks, 
adopting the axisymmetric semi-analytical 2D model developed by \citet{visser09b,visser11} and adjusted by \citet{harsono13}.
We briefly present features of the model; more details can be found in the original papers.
The model describes the temporal evolution of the density and velocity fields in a semi-analytical manner,
following the inside-out collapse model with the effect of rotation \citep{shu77,cassen81,terebey84} 
and the one-dimensional accretion disk model with the $\alpha$ viscosity prescription \citep{shakura73,lynden-bell74,visser10}.
The vertical density structure of the disk is set by the assumption of hydrostatic equilibrium, 
i.e., self-gravity of the disk is not considered.
The disk surface is defined as regions where the ram pressure of infalling gas from the envelope to the disk 
equals the thermal pressure of the disk.
The effect of magnetic fields is neglected except for the consideration of outflow cavities, which are added manually.
The motivation behind this addition is to explore the impact of outflow cavities on chemistry through enhanced temperatures and stellar FUV radiation field.
The position of the outflow cavity wall at a given time $t$ is 
\begin{align}
z = (0.191\,\,{\rm au}) \left(\frac{R}{1\,\,{\rm au}}\right)^{1.5}\left(\frac{t}{t_{\rm acc}}\right)^{-3},
\end{align}
where $R$ and $z$ are cylindrical coordinates \citep[see][for more details]{drozdovskaya14}.
$t_{\rm acc}$ is defined as $M_0/\dot{M}$, where $M_0$ is the initial core mass 
and $\dot{M}$ is the accretion rate of infalling gas onto the star-disk system \citep{shu77}.
The $t^{-3}$ dependence means that the opening angle of the cavity increases with time.
Inside the cavity, the gas density is set to be a constant 10$^{4}$ cm$^{-3}$. 

The dust temperature and stellar FUV (far-UV, 6-13.6 eV) radiation field are critical for the chemistry.
Given the density structure, they are calculated by solving the wavelength-dependent radiative transfer at each timestep with RADMC-3D\footnote{http://www.ita.uni-heidelberg.de/\~{}dullemond/software/radmc-3d/}.
The interstellar radiation field is neglected under the assumption that the core is embedded in an ambient molecular cloud.
The gas and dust temperatures are initially set to be 10 K throughout the core.
It should be noted that viscous heating, which is important for the temperature of the inner disk \citep{dalessio97,harsono15a} is not considered. 
Since the main focus of this paper is the delivery of water ice from a protostellar envelope to a disk rather than processing in a disk itself,
the omission does not affect our primary conclusions.
In low density and UV irradiated regions, the gas temperature can be higher than the dust temperature \citep[e.g.,][]{draine78}.
We calculate the temperature difference between gas and dust using an analytical prescription based on detailed 
thermo-chemical models by \citet[][]{bruderer12} (S. Bruderer 2015, private communication).  
The impact of the gas-temperature modification is not significant at least for water chemistry discussed in this paper.

Initially, the core has a power-law density distribution $\propto r^{-2}$, where $r$ is the distance from the center of the singular core, 
with an outer boundary of $\sim$7000 au and core mass of $M_0 = 1 M_{\odot}$.
The effective sound speed, which is a parameter to set the speed of the collapse expansion wave, is set to $c_s = 2.6 \times 10^{4}$ cm s$^{-1}$.
In our fiducial model\footnote{Our fiducial model corresponds to case 7 in \citet{visser09b} and `infall-dominated disk' case in \citet{drozdovskaya14}.}, 
a solid-body rotation rate of the core is assumed to be $\Omega = 10^{-13}$ s$^{-1}$, 
and thus the ratio of rotational energy to the gravitational energy is only $\sim$0.8 ($\Omega/10^{-13}$ s$^{-1}$) \%.
The adopted energy ratio is similar to the typical value inferred from observations of dense cores \citep{goodman93,caselli02}.
The $\alpha$-viscosity of the disk is set to $\alpha_{\rm vis} = 10^{-2}$ regardless of space and time.
The protostar is assumed to be born at $2\times10^4$ yr after the onset of collapse \citep[see][for the detailed discussion]{visser09b}. 
The model follows the physical evolution until the end of the main accretion phase, $t=t_{\rm acc}\sim2.5\times10^5$ yr,
when the gas accretion onto the star-disk system is almost complete ($>$99 \% of the total mass is in the star-disk system).

Figure \ref{fig:phys} shows the physical structure at $t = 0.5t_{\rm acc}$ and 0.9$t_{\rm acc}$ in our fiducial model.
The disk outer radius increases with time and finally reaches $\sim$280 au at $t=t_{\rm acc}$ (not shown).
The disk is denser and colder than the envelope material, but warmer than 20 K.
It is seen that the envelope regions close to the outflow cavity walls are subject to higher levels of stellar UV radiation,
while the disk is heavily shielded from UV irradiation.

\section{Chemical model}
\label{sec:chem}
Fluid parcels are traced in the physical model, and a gas-ice chemical simulation is performed 
along the streamlines \citep[e.g.,][]{drozdovskaya14}.
The chemical evolution along a total of $\sim$35,000 streamlines is calculated for our fiducial model 
in order to investigate the envelope-scale and disk-scale chemical structures with their time evolution.
The $\sim$35,000 fluid parcels are randomly distributed throughout the core at the onset of the collapse.
We adopt the rate equation method and the chemistry is described by a seven-phase model,
which is introduced as a natural extension of the three-phase model proposed by \citet{hasegawa93b}.
The three-phase model considers the gas phase, a surface of ice, and the bulk ice mantle, 
assuming that the bulk mantle has uniform chemical composition for simplicity.
Our seven-phase model takes into account a depth-dependent ice structure to some extent 
by considering the bulk mantle as five distinct phases at most.
We chose the seven-phase model, considering the balance between the computational time and the resolution of an ice layered structure.
A detailed description of the method is given in Appendix \ref{appendix:mph}.
As chemical processes, we consider gas-phase reactions, interaction between gas 
and grain/ice surface, and surface reactions.
We assume the top four monolayers of the ice are chemically active following \citet{vasyunin13}, i.e., the top four ice monolayers
are considered to be a surface.
The bulk ice mantle is assumed to be chemically inert in our fiducial model (i.e., $\req{P}{i}{m_j} = \req{L}{i}{m_j} = 0$ in Equation (\ref{eq:mph3})).

We use the same chemical network as in \citet{furuya15} except that species containing 
chlorine, phosphorus, or more than four carbon atoms were excluded.
The network is originally based on the gas-ice chemical reaction network of \citet{garrod06} supplemented by 
the high-temperature gas-phase reaction set from \citet{harada10}.
The network has been extended to include singly, doubly, and triply deuterated species \citep{aikawa12,furuya13}, 
and nuclear spin states of light species \citep{hugo09,hincelin14}.
In total, the network consists of 728 gaseous species, 305 icy species for each ice phase, and $\sim$85,000 reactions.

The binding energy of water on a surface is set to be 5700 K \citep{fraser01}, 
which corresponds to a sublimation temperature of 100-150 K for the relevant density range in our model.
Since the details of our treatment of the chemical processes are given in \citet{furuya15}, 
we present here only the treatment of chemistry induced by UV photons and the adopted initial molecular composition for the collapse model.
Unlike \citet{furuya15}, the modified rate method of \citet{garrod08}, which can take into account the competition between surface processes in the stochastic regime, 
is not used in this work in order to shorten the computational time.
We confirmed that this omission does not significantly affect the water chemistry discussed in the present paper, by running our chemical model 
with the modified rate method for several streamlines.
The potential rationales for the non-significant stochastic effect are that 
(i) a moderately high thermal hopping-to-desorption energy ratio of 0.5 is assumed in our model 
(also the surface diffusion through quantum tunneling is not allowed in our model), and 
(ii) our collapsing cores have a relatively high gas density ($>$10$^4$ cm$^{-3}$ for 10 K and orders of magnitude higher densities for warmer regions).
Both the slower surface diffusion and higher gas density (i.e., higher adsorption rates onto dust grains) tend to reduce the stochastic effect \citep[e.g.,][]{vasyunin09}.

\subsection{Chemistry induced by UV photons}
\subsubsection{Gas phase}
We consider chemistry induced by stellar and cosmic ray-induced UV photons.
In the physical model, the wavelength-dependent stellar UV flux is calculated at each point in the core.
We approximate the photodissociation/ionization rates by scaling the rates for the interstellar radiation field 
using the wavelength-integrated FUV flux at each point.
Self-shielding for H$_2$, HD, CO, and N$_2$ and mutual shielding by H$_2$ for HD, CO, and N$_2$ 
are taken into account \citep{draine96,visser09a,wolcott11,li13}.
The effective column densities of H$_2$ and other self-shielding species at each point are calculated 
following the method of \citet{visser11} (see their Section 3.1).
Photorates for cosmic ray-induced UV are calculated following \citet{gredel89}.

\subsubsection{Solid phase}
Photodissociation of water ice can lead to several outcomes including desorption of water and its photofragments 
into the gas phase \citep{andersson08}.
The total photodissociation rate (cm$^{-3}$ s$^{-1}$) of water ice on an icy grain surface is 
calculated similar to \citet{furuya15}:
\begin{align}
R^{\rm tot}_{{\rm ph}, \,i} &= f_{{\rm abs}, \,i} \pi a^2 n_{\rm gr} F_{\rm UV}, \label{eq:photodiss} \\ 
f_{{\rm abs}, \,i} &= \theta_i P_{{\rm abs}, \,i} \times {\rm min}(N_{\rm layer}, \,\, 4), \label{eq:photodiss2}
\end{align}
where $n_{\rm gr}$ is the number density of dust grains per unit gas volume, 
$a$ is the radius of dust grains (10$^{-5}$ cm), 
$f_{{\rm abs}, \,i}$ is the fraction of the incident photons absorbed by species $i$ (i.e., water ice here), and 
$F_{\rm UV}$ is the flux of stellar or cosmic-ray induced UV photons.
$\theta_i$ is the surface coverage of species $i$, which is defined as a fractional abundance of  species $i$ in the top four monolayers 
(i.e., on the surface; $\theta_i = \req{n}{i}{s}/\sum_{k} \req{n}{k}{s}$, where $\req{n}{i}{s}$ is the number density of species $i$ on the surface).
$P_{{\rm abs}, \,i}$ is the probability of photoabsorption by one monolayer of pure water ice,
which is calculated by convolving the wavelength-dependent photoabsorption cross sections of 
water ice \citep{mason06} with the emission spectrum of the interstellar radiation
field \citep{draine78} and with that of the cosmic ray-induced radiation field \citep{gredel89}.
$N_{\rm layer}$ is the number of monolayers of ice in total, and the term ${\rm min}(N_{\rm layer}, \,\, 4)$ 
in $f_{{\rm abs}, \,i}$ corresponds to the absorption by up to four outermost monolayers.
\citet{arasa15} studied the possible outcomes of H$_2$O, HDO, and D$_2$O photodissociation 
in H$_2$O ice and derived their probabilities per dissociation ($b_j$), using molecular dynamics simulations.
With their results, the rate of each outcome $j$ is calculated by $b_jR^{\rm tot}_{{\rm ph}, \,i}$.
The photodesorption yield of water (desorbed as OX or X$_2$O, where X is H or D) per incident UV photon is $\sim3\times10^{-4}$ in our model.

For the other icy species, the photoabsorption probability is calculated by $P_{{\rm abs}, \,{\rm H_2O}} \times k^0_{i}/k^0_{\rm H_2O} $,
where $k^0_{i}$ is the photodissociation rate in the gas phase for the interstellar radiation field.
It is assumed that all photofragments remain on the surface for simplicity.
Instead the photodesorption rates are calculated by Equations (\ref{eq:photodiss}) and (\ref{eq:photodiss2}), 
but $P_{{\rm abs}, \,i}$ and ${\rm min}(N_{\rm layer}, \,\, 4)$ are replaced 
by the experimentally derived photodesorption yield and ${\rm min}(N_{\rm layer}/4, \,\, 1)$, respectively \citep[e.g.,][]{bertin12,fayolle13}.
For species for which the yield is not available in the literature, we assume a yield of 10$^{-3}$ per incident photon.

UV photons can penetrate deep into the ice and dissociate molecules in the bulk ice mantle as well as those on the surface.
In our fiducial model, it is implicitly assumed that the photofragments in the ice mantle recombine immediately.
The effect of this assumption is discussed in Section \ref{sec:bulkice}.

\subsection{Initial abundances}
\label{sec:init}
In order to set the molecular abundances at the onset of collapse, we reran the model of 
molecular cloud formation from diffuse \mbox{\ion{H}{i}} dominated clouds for $8\times10^6$ yr \citep{bergin04,hassel10,furuya15}, 
followed by a further molecular evolution for $3\times10^5$ yr 
under prestellar core conditions; the gas density, temperature, 
and the visual extinction are set to $2\times 10^4$ cm$^{-3}$, 10 K, and 10 mag, respectively.
The cosmic ray ionization rate of H$_2$ is set to be $5\times10^{-17}$ s$^{-1}$ \citep{dalgarno06}.
The elemental abundance of deuterium with respect to hydrogen is set to $\dhr{elem}=1.5\times10^{-5}$ \citep{linsky03}.

\begin{figure}
\resizebox{\hsize}{!}{\includegraphics{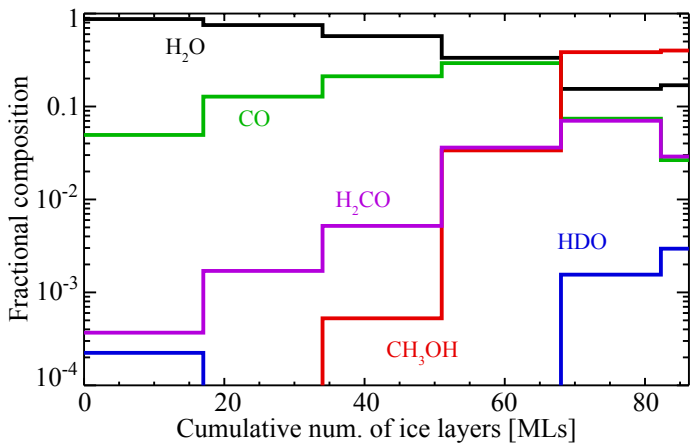}}
\caption{Initial layered ice structure at the onset of collapse. The ice consists of 86 MLs in total.
The ice surface (top 4 MLs) and five distinct mantle phases have different chemical compositions.}
\label{fig:init_chem}
\end{figure}

Figure \ref{fig:init_chem} shows the initial ice layered structure for the collapse model.
In the inner ice layers (i.e., ice layers formed in the early evolution; $<$50 MLs) H$_2$O is the dominant constituent, while in the outer layers
the fraction of CO and its hydrogenated molecules become significant.
The trend is qualitatively consistent with infrared observations of the ISM ice \citep[][]{pontoppidan06,oberg11}.
It is also seen that the fraction of deuterated water increases toward the outer layers as discussed in Section \ref{sec:intro}.
The total H$_2$O ice abundance with respect to gaseous hydrogen nuclei is $1.1\times10^{-4}$.
The $\hdo$ ratio and the $\ddo$ ratio in the entire ice mantle are $8.7\times10^{-4}$ and $4.9\times10^{-3}$, respectively,
which reproduce the observationally derived values toward the Class 0 protostar NGC 1333-IRAS 2A within a factor of 2 \citep[][]{coutens14}.
The agreement between the modeled and observational values justifies our initial layered ice structure for the collapse model \citep[see][for further discussion]{furuya16b}.  
 
In Table \ref{table:parameters}, we summarize the parameters considered in this paper.
The impact of some of them is discussed in Section \ref{sec:parameter}.

\begin{table}
\caption{Summary of adopted parameters.}
\label{table:parameters}
\centering
\begin{tabular}{lc}
\hline\hline
Parameters &  Values\\
\hline
$M_{0}$ ($M_{\odot}$)  & 1  \\
$\Omega$ (s$^{-1}$) & 10$^{-14}$ - $\boldsymbol{10^{-13}}$ \\
c$_s$ (cm s$^{-1}$) & $2.6 \times 10^{4}$ \\
$\alpha_{\rm vis}$  & 10$^{-2}$  \\
Bulk ice chemistry & On - {\bf Off} \\
\hline
\end{tabular}
\tablefoot{Values used in our fiducial model are shown in bold letters.}
\end{table}
 
\section{Results from the fiducial model}
\label{sec:result}
\subsection{Infalling envelope}

\begin{figure}
\resizebox{\hsize}{!}{\includegraphics{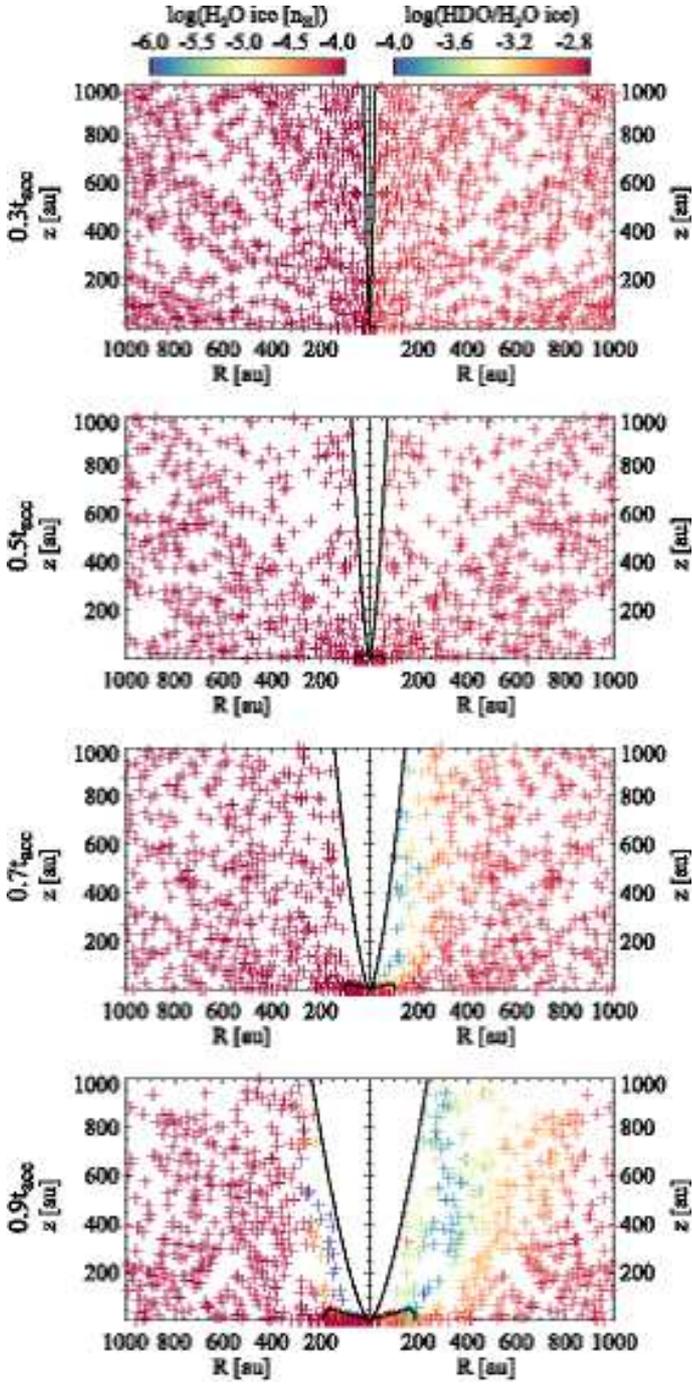}}
\caption{Spatial distributions of fluid parcels on a 1000 au scale at $t = 0.3t_{\rm acc}$, 0.5$t_{\rm acc}$, 0.7$t_{\rm acc}$, and
0.9$t_{\rm acc}$ (from top to bottom) in our fiducial model. 
Left: H$_2$O ice abundance with a logarithmic scale.
Right: ice $\hdo$ ratio with a logarithmic scale.
The black solid lines at each panel depict the disk surface and the outflow cavity wall.
See Figure \ref{fig:chem_zoom} for the zoom-in view on the disk.}
\label{fig:chem}
\end{figure}


\begin{figure}
\resizebox{\hsize}{!}{\includegraphics{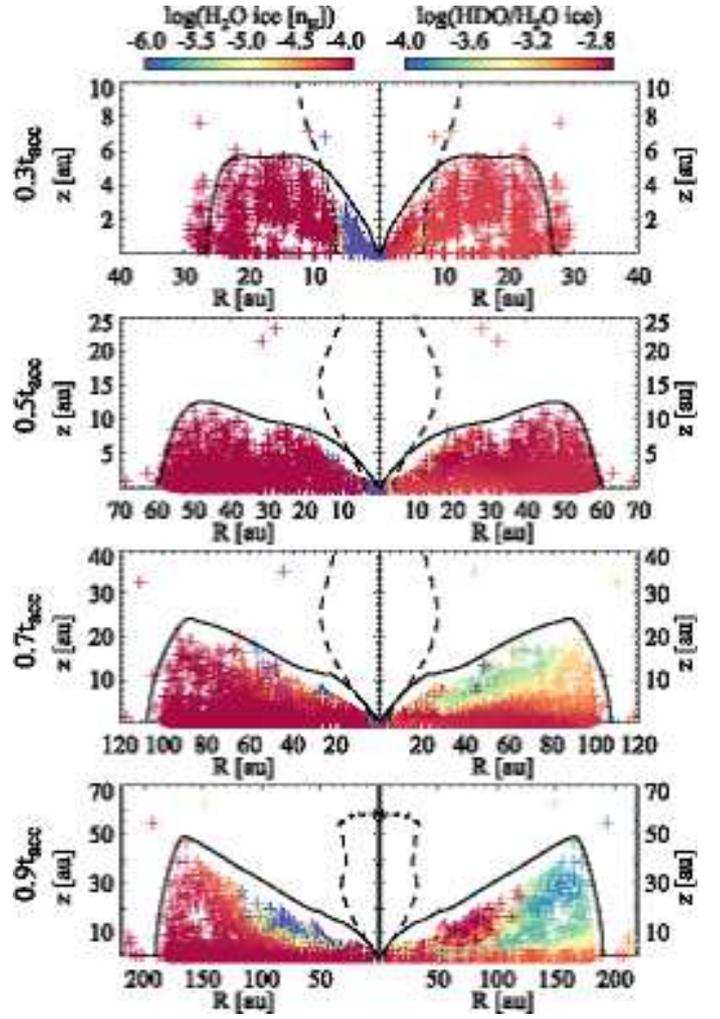}}
\caption{Zoom-in view of Figure \ref{fig:chem} on the disk scale. 
Note the different spatial scales among the panels.
Water snow lines, where the rate of adsorption onto dust grains and that of thermal desorption are balanced, are depicted by dashed lines.
The outflow cavity walls are not shown in this figure.}
\label{fig:chem_zoom}
\end{figure}

Given the complexity of our physical and chemical models, we present the global picture of water ice chemistry 
rather than discuss the detailed chemical evolution of individual fluid parcels.

Figure \ref{fig:chem} shows spatial distributions of the fluid parcels on a 1000 au scale  
at $t = 0.3t_{\rm acc}$, $0.5t_{\rm acc}$, $0.7t_{\rm acc}$, and $0.9t_{\rm acc}$.
At $t = 0.3t_{\rm acc}$ for example,  the star-disk system has $\sim$30 \% of the total core mass, while the envelope has $\sim$70 \% of the total mass.
Each fluid parcel experiences different physical conditions in the infalling protostellar envelope \citep[see Figure 7 in][for examples]{drozdovskaya14}.
Reflecting different histories, the H$_2$O ice abundance (left panels) and the $\hdo$ ice ratio (right panels) vary among the fluid parcels. 
It is seen that the H$_2$O ice abundance is almost unaltered in the infalling envelope except for the regions close to the outflow cavity wall, 
in which H$_2$O ice is largely destroyed by stellar UV photons.
The size of the regions where H$_2$O ice is lost increases with time as the opening angle of the cavity increases and the density of the envelope decreases,
both of which allow the deeper penetration of the stellar UV photons into the envelope. 
Note that the regions with a high dust temperature ($>$100 K), where water ice sublimates, are small ($<$ several of tens au, depending on time) 
in the fiducial physical model.
Thus thermal desorption is not important for the water abundance structure on a 1000 au scale.
Our results are consistent with the earlier findings by \citet{visser11},
i.e., the majority of H$_2$O is delivered to the disk as ice without significant UV processing and sublimation, 
while some H$_2$O is lost in the protostellar envelope prior to disk entry.

The $\hdo$ ratio is found to be more sensitive to the stellar UV photons in the envelope (Figure \ref{fig:chem}, right panels).
The $\hdo$ ratio is lower than the initial value ($\sim$10$^{-3}$) even in regions where the H$_2$O ice abundance is 
almost unaltered (see Section \ref{sec:details} for more quantitative discussions).
This is a consequence of the layered structure of the ice; 
most HDO is present in the upper layers of ice mantles, which are destroyed via photodissociation and thermal desorption of the photofragments
prior to the destruction of the lower ice layers, where most H$_2$O is present (see Figure \ref{fig:init_chem}).
The selective loss of HDO ice leads to the decrease in the $\hdo$ ratio, depending on streamline. 
In our fiducial model, water ice in the protostellar envelope has a variation in the $\hdo$ ice ratio 
ranging from $\sim$10$^{-4}$ to $\sim$10$^{-3}$.
We confirmed that the selective loss of HDO ice does not occur if we assume a well-mixed ice at the onset of collapse.

In summary, there are three cases for water ice chemistry in the infalling envelope, depending on the degree of 
UV processing experienced along the streamline:
(i) the prestellar ice is almost preserved, and both the H$_2$O ice abundance and the $\hdo$ ice ratio are almost unaltered during the collapse,
(ii) only the upper layers of the prestellar ice are lost and then the H$_2$O ice abundance is almost unaltered, 
while the $\hdo$ ice ratio is lowered, and 
(iii) the entire H$_2$O ice reservoir is largely lost (and partly reformed). 
The importance of UV processing increases with time, which can lead to a time-dependent disk composition.
We note that in case (ii), not only HDO ice but also other icy molecules which are formed later than H$_2$O ice in the prestellar phase, such as 
regular/deuterated formaldehyde and methanol ices, are selectively lost with respect to H$_2$O ice (see Section \ref{sec:organics}).

\subsection{Disk}
\label{sec:disk}

Figure \ref{fig:chem_zoom} shows the H$_2$O ice abundance (left panels) and the $\hdo$ ice ratio (right panels) 
on the spatial scale of the forming disk at the selected times.
Reflecting the variation of the composition of the accreting materials, the disk composition varies with time.
It is also seen that the disk composition at any given time is not spatially uniform, especially at later times. 

For more quantitative discussions, we calculate the vertically averaged H$_2$O ice abundance and the $\hdo$ column density ratio 
as functions of disk radius.
We interpolated the molecular compositions of the fluid parcels to the cylindrical coordinates used in the physical simulation,
adopting bilinear interpolation.
From the gridded data, the vertically averaged abundance and the column density ratio for $z/R < 0.2$ at each radius are calculated, 
because the spatial distributions of the fluid parcels are sparser at $z/R \gtrsim 0.2$.
We confirmed that at least 55-80 \% of the gas mass at each radius lies below $z/R = 0.2$, depending on time, 
and that a change of the threshold $z/R$ (0.15 or 0.25) does not affect the resultant abundance and column density ratio significantly.
A small number of the fluid parcels have very high $\hdo$ ice ratios of 10$^{-2}$-10$^{-1}$ 
(30 parcels out of $\sim$12,000 parcels at $t=t_{\rm acc}$ for example).
Those parcels were not considered in the above calculation, because their inclusion with the interpolation 
makes the radial profile of the $\hdo$ ratio very peaky in some regions ($R$ = 50-130 au at $t=t_{\rm acc}$ for example).
The importance of such super deuterated water for the disk composition is unclear from our simulations.

\begin{figure}
\resizebox{\hsize}{!}{\includegraphics{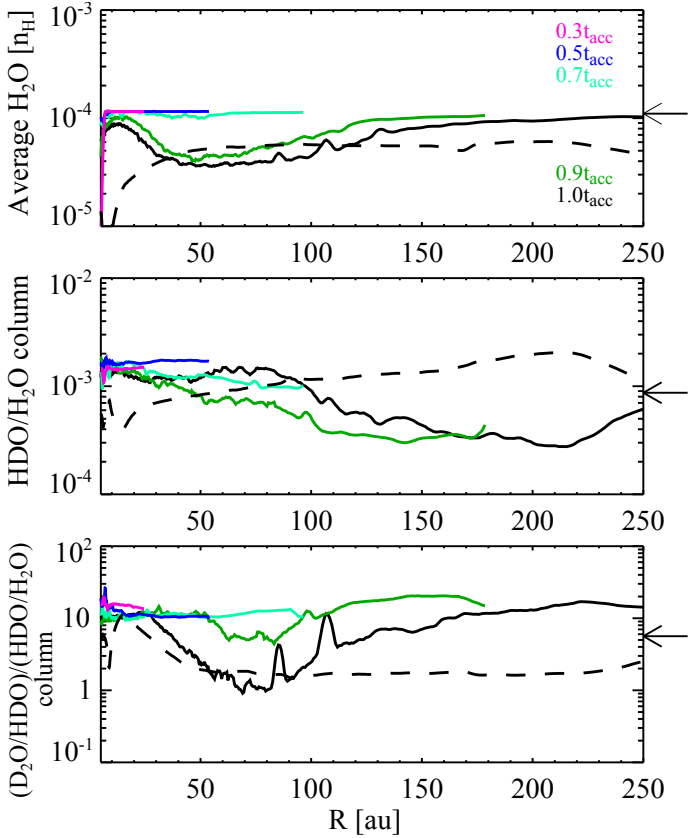}}
\caption{Vertically averaged H$_2$O ice abundance (top), $\hdo$ ice column density ratio (middle), 
and ice column density ratio of $\ddo$ to $\hdo$ (bottom) in the forming disk 
as a function of radius at selected times in our fiducial model, represented by different colors (solid lines). 
Arrows on the right-hand margin indicate the values at the onset of collapse.
Note that the disk outer radius increases with time.
The black dashed lines represent the prediction assuming in-situ formation of 
water ice in the disk structure given at $t=t_{\rm acc}$.
}
\label{fig:column}
\end{figure}

Figure \ref{fig:column} shows the vertically averaged H$_2$O ice abundance (top) and the $\hdo$ column density ratio (middle)
as functions of radius at selected times.
At $t < 0.7t_{\rm acc}$, the averaged H$_2$O ice abundance is very similar to the initial (precollapse) value of 10$^{-4}$.
At later times, the H$_2$O ice abundance becomes lower than 10$^{-4}$ at  30 au $\lesssim R \lesssim$ 100 au 
by up to a factor of 3, owing to the accretion of H$_2$O ice-poor material onto the disk.

The $\hdo$ ice ratio is almost constant throughout the disk at $t < 0.7t_{\rm acc}$, and the ratio is higher 
than the precollapse value ($9\times10^{-4}$) by a factor of up to 2,
owing to the additional formation of a small amount (H$_2$O abundance of $\sim$10$^{-6}$) of highly fractionated ($\hdo > 10^{-2}$) water ice during the collapse.
At later times, the $\hdo$ ice ratio tends to decrease with radius, 
owing to the accretion of material with a low $\hdo$ ice ratio ($\sim$10$^{-4}$) onto the outer disk.
This radial trend contrasts with the trend expected from the disk thermal structure;
as the gas temperature decreases with radius, the level of deuterium fractionation is expected to increase with radius.
For confirmation we ran the chemical model on the disk structure given at $t=t_{\rm acc}$ for a selection of $\sim$200 spatial points,
assuming that chemical species are initially in atomic form except for hydrogen and deuterium
which are locked up in H$_2$ and HD, respectively.
We assumed that the ortho-to-para ratio of H$_2$ is locally thermalized (i.e., $\op{H_2} = 9\exp(-170/T_{\rm gas})$) in this simulation for simplicity. 
After $2.5\times10^5$ yr (=$t_{\rm acc}$), the vertically averaged abundance of H$_2$O ice reaches $\sim5\times10^{-5}$ 
at $R \gtrsim 30$ au, inside which the production of H$_2$O ice is less efficient due to the higher dust temperatures ($\gtrsim$30 K) 
even in the midplane. 
The $\hdo$ ice column density ratio increases with radius as expected (dashed line in Figure \ref{fig:column}).
The qualitative difference between the radial profile of the $\hdo$ ratio in our fiducial model and 
that predicted from the disk thermal structure demonstrates the importance of the dynamical evolution 
and the ice layered structure, with the latter mostly established in the precollapse phase.

\begin{figure}
\resizebox{\hsize}{!}{\includegraphics{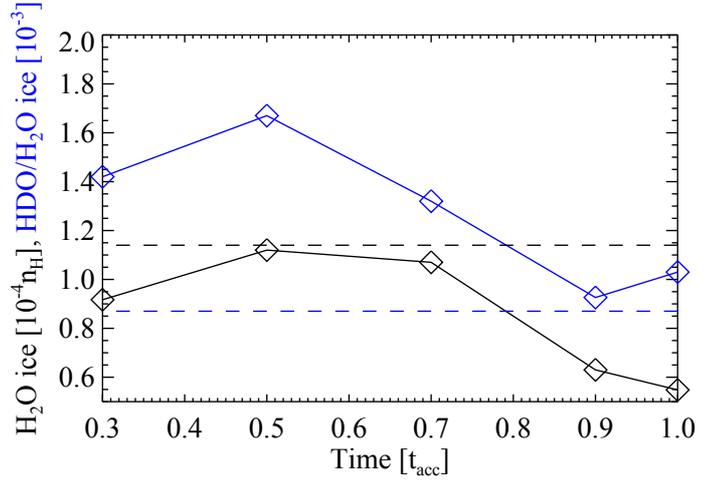}}
\caption{Bulk disk-averaged H$_2$O ice abundance (black) and $\hdo$ ice ratio (blue) as functions of time in the fiducial model.
The horizontal dashed lines indicate their values at the onset of collapse.
}
\label{fig:total}
\end{figure}
From the radial profiles of the H$_2$O ice and HDO ice column densities, 
the average H$_2$O ice abundance and $\hdo$ ice ratio in the whole disk are calculated, assuming axisymmetry.
Figure \ref{fig:total} shows the bulk disk-averaged H$_2$O ice abundance and $\hdo$ ice ratio as functions of time.
The deviations from the precollapse values are factors of $\sim$2 at most, regardless of time.
Therefore, when averaged over the whole disk, the forming disk contains an H$_2$O abundance and a $\hdo$ ratio 
similar to those at the onset of collapse in our model,
while the local vertically averaged disk compositions show larger differences, especially at late times (Figure \ref{fig:column}).

\subsection{Details of water and deuterium chemistry}
\label{sec:details}
The above subsections have shown that water chemistry in our model is dominated by processes induced by stellar UV photons in the envelope.
For a more detailed analysis, we define normalized fluence\footnote{Fluence is the time integral of flux with a unit of cm$^{-2}$ in cgs.}
for each fluid parcel along its streamline, $\overline{Ft}$, as
\begin{align}
\overline{Ft} = \int_{0}^{t_{\rm acc}} [F_{\ast}(\bm{l}(t)) + F_{\rm CRUV}] dt / \int_{0}^{t_{\rm acc}} F_{\rm CRUV} dt, \label{eq:ft}
\end{align}
where $F_{\ast}$ and $F_{\rm CRUV}$ are the local flux by the stellar and cosmic-ray induced UV photons, respectively, and 
$\bm{l}(t)$ is the position vector of a fluid parcel in cylindrical coordinates at a given time $t$.
Since the cosmic ray-induced UV photons set the minimum of the UV radiation field in our model, the normalized fluence can be used 
as a measure of the importance of the stellar UV radiation.
In our model, $F_{\rm CRUV}$ is constant, 10$^4$ photons cm$^{-2}$ s$^{-1}$, for simplicity, neglecting an attenuation effect of cosmic rays 
for large column densities \citep{umebayashi81}.
It is worth mentioning again that the disk is heavily shielded from the stellar UV irradiation (Figure \ref{fig:phys}).
$\overline{Ft} = 1$, i.e, when the cosmic ray-induced UV dominates over the stellar UV, corresponds to 
$\sim2\times10^{7}$ incident UV photons per one dust grain up to $t_{\rm acc}$ in our model.

\begin{figure}
\resizebox{\hsize}{!}{\includegraphics{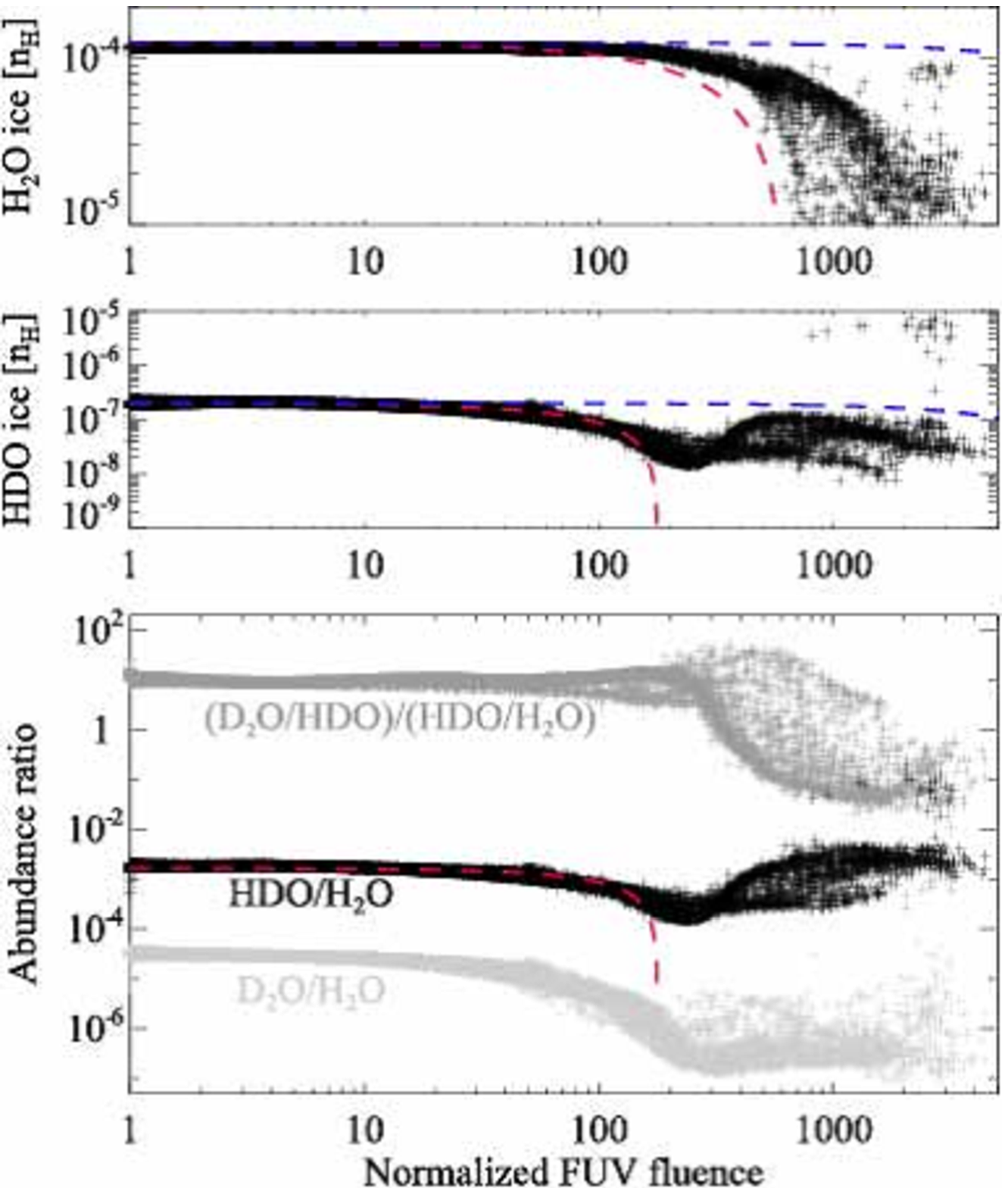}}
\caption{H$_2$O ice (top)  and HDO ice (middle) abundances in the fluid parcels that are located in the disk at $t=t_{\rm acc}$ 
as functions of the normalized UV fluence.
The bottom panel shows $\hdo$ ice ratio (black), D$_2$O/H$_2$O ice ratio (light gray), and the ratio of $\ddo$ to $\hdo$ ice ratio (gray).
Only the fluid parcels in which the H$_2$O ice abundance is larger than 10$^{-5}$ are plotted.
The blue and red dashed lines depict the expected H$_2$O (or HDO) ice abundance when only photodesorption is 
an allowed chemical process and when only photodissociation/photodesorption are included, respectively.
See the main text for more details.
}
\label{fig:uv_ab_case7}
\end{figure}

Figure \ref{fig:uv_ab_case7} shows the abundances of  H$_2$O ice (top) and HDO ice (middle), and the $\hdo$ ice ratio (bottom) 
 as functions of $\overline{Ft}$ in the fluid parcels which are located in the disk at $t=t_{\rm acc}$.
It is clear that the cosmic ray-induced UV has a negligible impact on the water ice abundances, while the stellar UV dominates the water ice chemistry.
An anti-correlation between the UV fluence and the H$_2$O ice abundance is seen especially at $\overline{Ft} \gtrsim 500$.
The dashed blue and red lines in the figure represent the expected water ice abundances as functions of $\overline{Ft}$ 
when only photodesorption is an allowed chemical process and when only photodissociation/photodesorption are allowed, respectively:
\begin{align}
\ab{HXO_{ice}}(\overline{Ft}) = \ab{HXO_{ice}}^0 - (\Sigma_{j} b_j) \times \overline{R^{\rm tot}_{{\rm ph, \, HXO_{ice}}}} t_{\rm acc}, \label{eq:nh2o}
\end{align}
where $\ab{HXO_{ice}}^0$ is the abundance of HXO ice, where X is H or D, in parcels with $\overline{Ft} \sim 1$.
$\overline{R^{\rm tot}_{{\rm ph, \, HXO_{ice}}}}$ is similar to Equation (\ref{eq:photodiss}) but $n_{\rm gr}$ and $F_{\rm UV}$ are replaced 
by the abundance of dust grains and the time-averaged flux along a streamline, $\overline{Ft} \times F_{\rm CRUV}$, respectively.
On the evaluation of Equation (\ref{eq:nh2o}), $\theta_{\rm H_2O}$ was set to be 0.35 (i.e., the rest of the ice mantle consists of other species), 
which comes from the average fraction of H$_2$O in the topmost $\sim$50 MLs of the ice mantle at the onset of collapse (cf. Figure \ref{fig:init_chem}).
That is smaller than the average fraction of H$_2$O in the entire ice mantle, $\sim$0.5, but larger than the surface coverage of H$_2$O ice, $\sim$0.2.
$b_j$ is the branching ratio of each outcome $j$ of H$_2$O ice photodissociation \citep{arasa15}.
It is clear that the photodissociation of H$_2$O ice is much more important for H$_2$O ice destruction than photodesorption.
Given the warm temperature of the protostellar envelope ($\gtrsim$20-30 K, see Figure \ref{fig:phys}), 
which reduces the probability of hydrogenation on the grain surface upon H atom adsorption, 
the reformation of H$_2$O ice is not efficient enough to compensate for its photodissociation.
However, considering the numerical data are (well) above the red line, the reformation of H$_2$O ice is not negligible. 
The scatter in the H$_2$O ice abundance at the high fluence regime likely comes from the difference in the dust temperature
when water ice is (partly) reformed.
The reformation of H$_2$O ice via OH$_{\rm ice}$ + H$_{\rm ice}$ competes with the formation of  CO$_2$ ice via OH$_{\rm ice}$ + CO$_{\rm ice}$, 
the latter of which becomes more favorable with increasing dust temperature.
At higher dust temperatures ($\gtrsim$40-50 K), where surface chemistry is inefficient, 
gaseous O$_2$ is the dominant product of water ice destruction, which forms via the neutral-neutral reaction in the gas phase, OH + O \citep[see also][]{drozdovskaya16,taquet16}.

The HDO ice abundance and the $\hdo$ ice ratio show more complex behavior: 
they start to drop at $\overline{Ft} \sim 100$,
while in the higher fluence regime, where H$_2$O ice starts to be largely destroyed and partly reformed, 
they tend to stay constant or increase with the fluence.
The former indicates again that the HDO ice abundance is more sensitive to the photochemistry than that of H$_2$O ice; 
the upper ice layers where most HDO is present are lost prior to the lower ice layers where most H$_2$O is present.
The red line in the middle panel shows Equation (\ref{eq:nh2o}) for HDO with $\theta_{\rm HDO}$ of $2\times10^{-3}$ (cf. Figure \ref{fig:init_chem}).
The line reproduces the numerical data at $\overline{Ft} \lesssim 200$.
The behavior of the HDO ice abundance in the higher fluence regime indicates that the level of deuterium fractionation 
in reformed water ice is similar to or higher than that in the original water ice (i.e., water ice before the destruction).
This implies that using the $\hdo$ ratio as a probe of the prestellar inheritance of H$_2$O is limited, 
because the $\hdo$ ratio does not allow distinction between the original and reformed ice.
Alternatively, in the bottom panel of Figure \ref{fig:uv_ab_case7}, it is seen that the ratio of $\ddo$ to $\hdo$ is different between the original and reformed water ice.
This implies that the ratio of $\ddo$ to $\hdo$ is a better probe of the prestellar inheritance of H$_2$O than solely the $\hdo$ ratio.
We discuss this probe of the prestellar inheritance in more detail in Section \ref{sec:inheritance}.

The rationale for the high $\hdo$ ratio in the reformed water ice is twofold;
(i)  the photofragments are highly enriched in deuterium as the original ice was highly enriched in deuterium, 
(ii) it takes some time for the photofragments to reach the equilibrium level (i.e., low level) of deuterium fractionation 
at the warm temperatures of the envelope via gas-phase ion-neutral chemistry.
Then, if the timescale of ice reformation is shorter than the relaxation timescale of deuterium fractionation, 
the fractionation in the photofragments (i.e, the original ice) is transferred to the reformed ice.
This is the case in our model.
Additionally there are pathways of deuterium fractionation in the gas phase which can work even at warm gas temperatures, 
e.g., OH + D $\rightarrow$ OD + H + 810 K \citep{millar89,atahan05}.
OD in the gas phase can freeze out onto (icy) dust grains, followed by the formation of HDO ice via OD$_{\rm ice}$ + H$_{\rm ice}$.


\section{Parameter dependencies}
\label{sec:parameter}
\subsection{Effect of bulk ice chemistry}
\label{sec:bulkice}
\begin{figure}
\resizebox{\hsize}{!}{\includegraphics{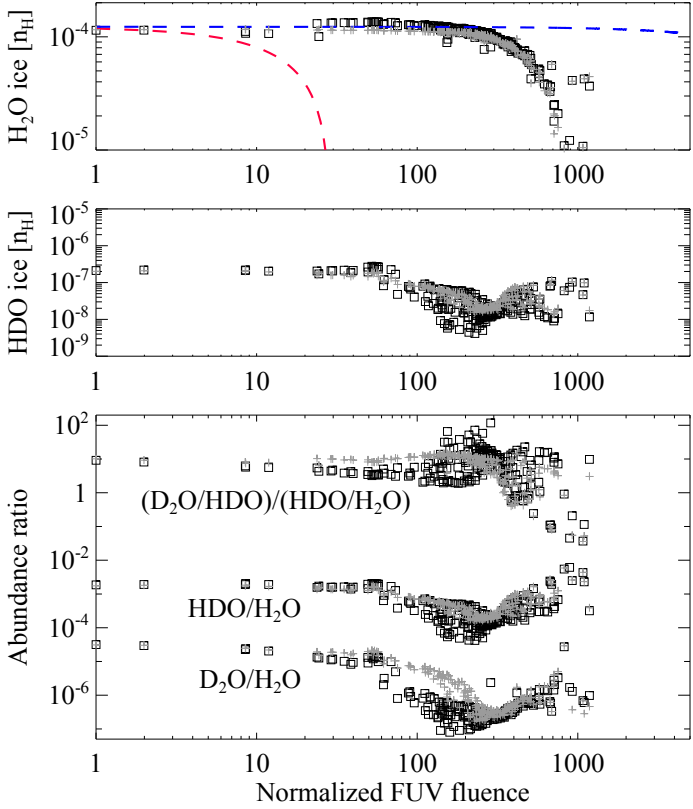}}
\caption{Similar to Figure \ref{fig:uv_ab_case7}, but for the model with the bulk ice chemistry (shown by square).
Gray crosses show the results from our fiducial model for comparisons.
}
\label{fig:uv_ab_bulk}
\end{figure}

In our fiducial model it is assumed that the top four monolayers of ice are chemically active, while the rest of the bulk mantle is inert.
However UV photons can penetrate deeper into the ice ($>$100 MLs) and dissociate icy molecules.
To check the impact of this assumption on the inert bulk mantle, we performed additional simulations in which bulk ice chemistry 
(including two-body reactions and photodissociation) in each mantle phase is taken into account 
(i.e., $\req{P}{i}{m_j} \ne 0$ and $\req{L}{i}{m_j} \ne 0$ in Equation (\ref{eq:mph3})).
For two-body reactions, we assume that species thermally diffuse in ice and react with each other when they meet 
analogously to grain-surface reactions \citep{garrod13}.
We only allow reactions between species within the same mantle phases, 
assuming that vertical motion of species is limited ($\lesssim$8 MLs, as each mantle phase has at most $\sim$17 MLs) 
and the longer-range vertical diffusion is not important.  
We set the energy barrier against thermal diffusion in ice to be 1.2$E_{\rm des}$, where $E_{\rm des}$ is the binding energy on a surface.
For photodissociation, we use a similar formula to Equation (\ref{eq:photodiss}), but additionally consider UV attenuation in ice 
by evaluating the photon absorption probability in each mantle phase in order from upper to lower mantle phases.
The effect of the UV attenuation is minor, as ice with a thickness of $\lesssim$100 MLs is optically thin against UV photons \citep{andersson08}.
For simplicity, we assume that all photofragments remain trapped in ice (i.e., the possibilities of reformation and desorption are not considered).
Since the inclusion of the bulk ice chemistry increases the runtime of the chemical model considerably, we ran the model only for 
a selection of $\sim$200 fluid parcels, which are located in the disk at $t=t_{\rm acc}$.

Figure \ref{fig:uv_ab_bulk} shows abundances of  H$_2$O ice (top) and HDO ice (middle), the $\hdo$ ice ratio, and the ratio of 
$\ddo$ to $\hdo$ (bottom) as functions of $\overline{Ft}$ in the selected $\sim$200 fluid parcels.
We confirm that the bulk ice chemistry does not affect our qualitative chemical results, 
while it can affect the deuterated water ice abundances by a factor of several or less, depending on the streamline.
The H$_2$O ice abundance is similar between the two models; reformation of H$_2$O ice via two body reactions (OH + H and OH + H$_2$) 
in the ice mantle compensates for the destruction by photodissociation in our model with the bulk ice chemistry.
The red line in the top panel shows the expected H$_2$O ice abundance when only photodissociation/photodesorption in the entire ice mantle is included.
The significant deviation of the numerical data from the red line indicates the importance of the reformation of water ice in the bulk mantle.
We note that for the assumed energy barrier (600 K), 
the timescales of thermal diffusion for H atom and H$_2$ in ice are longer than 1 Myr at 10 K.
The warm temperature of the protostellar envelope promotes the diffusion of such light species {\it inside ice} and the reformation of water in our model.
This contrasts with our claim in Section \ref{sec:details} that the warm temperature of the protostellar envelope 
slows down the reformation of water ice {\it on an icy grain surface}, because of the efficient thermal desorption of the light species at the ice surface.

\subsection{Effect of rotation rate of a prestellar core}
\label{sec:rotation}
The rotation rate of the initial prestellar core is one of the critical parameters for the formation and evolution of circumstellar disks.
In our fiducial physical model, the initial solid-body rotation rate of the core is assumed to be $10^{-13}$ s$^{-1}$.
In this subsection, we present the model with a lower rotation rate of 10$^{-14}$ s$^{-1}$
and discuss its impact on the water ice chemistry\footnote{
The model corresponds to case 3 in \citet{visser09b} and `spread-dominated disk' case in \citet{drozdovskaya14}}.
The other physical and chemical parameters are the same as in our fiducial model.
$t_{\rm acc}$ is the same as in our fiducial model.

\begin{figure}
\resizebox{\hsize}{!}{\includegraphics{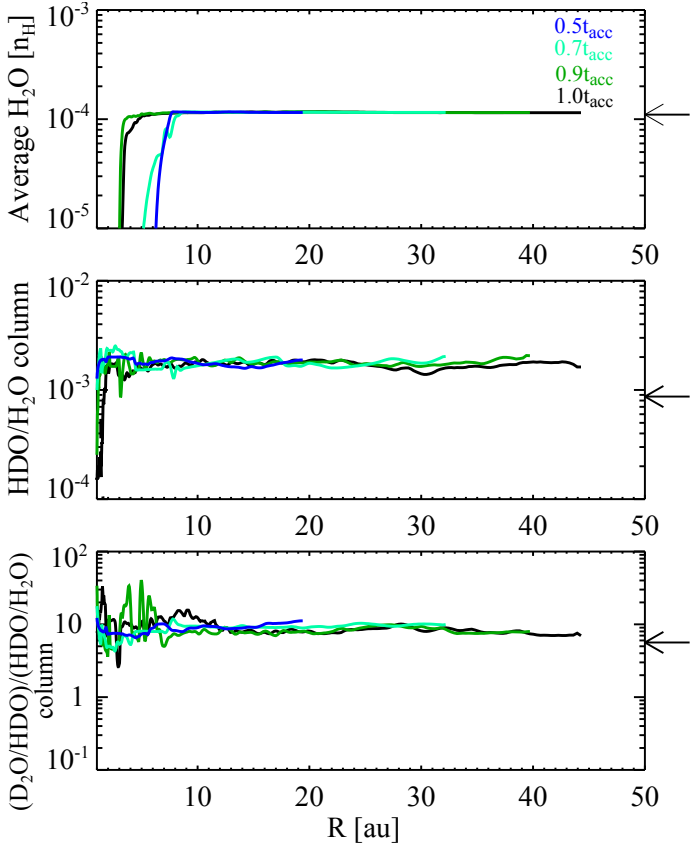}}
\caption{Similar to Figure \ref{fig:column}, but for the model with $\Omega=10^{-14}$ s$^{-1}$.
The results at $t = 0.3 t_{\rm acc}$ are not shown, because most of the disk has higher temperature than the water ice sublimation at the time.
}
\label{fig:column_c3}
\end{figure}

We confirmed that the spatial distributions of the H$_2$O ice abundance and the $\hdo$ ice ratio on a 1000 au scale are similar to those in the fiducial model;
the envelope material has variations in the H$_2$O ice abundance and the $\hdo$ ice ratio depending on the degree of processing by stellar UV photons 
(see Appendix \ref{appendix:rotation} for figures).
The variations in the envelope, however, are not directly transferred to the disk in the model with the lower rotation rate.
This contrasts with the fiducial model.
Figure \ref{fig:column_c3} shows the vertically averaged H$_2$O ice abundance and the $\hdo$ ice column density ratio in the disk 
in the lower rotational core model as functions of radius.
The outer disk radius at a given time is smaller than that in the fiducial model.
The sharp drop of the H$_2$O ice abundance in the inner regions is due to sublimation of water ice rather than the effect of UV photons.
Both H$_2$O ice abundance and $\hdo$ ice ratio are almost constant and independent of radius and time, 
but the $\hdo$ ratio is enhanced compared to the precollapse value by a factor of two.
The enhancement is caused by the additional formation of a small amount of highly deuterated water ice during the collapse as mentioned in Section \ref{sec:disk} for the fiducial model.
Because of the lower rotation rate, the envelope material which is exposed to the stellar UV radiation mostly 
accretes at small spatial scales (i.e., becomes part of the star) or enters into the outflow cavity and does not significantly contribute to the disk formation.
Note that the material which is significantly processed by the stellar UV radiation in the envelope tends to have lower specific angular momentum 
than that which is not significantly processed, since the outflow direction and the rotational axis are aligned.

In summary, the initial rotational rate does not affect the water ice chemistry significantly on 1000 au and larger spatial scales,
while it can affect the composition of forming disks as the fate of envelope material (falls onto stars, stays in disks, etc.) is largely decided by its initial specific angular momentum.
As a cautionary note, the temporal evolution of the star-disk-envelope system and the outflow cavity are not treated in a self-consistent way in our model.
Our model also does not consider the effect of magnetic fields explicitly, which may efficiently transfer angular momentum from disks to envelopes \citep[magnetic braking; e.g.,][]{basu94,yen15}. 
More sophisticated 3D non-ideal magnetohydrodynamics simulations of protostar/disk formation with chemistry are needed to fully confirm the above claim.

\section{Discussion}
\label{sec:discussion}
\subsection{Water deuteration as a probe of the prestellar inheritance of H$_2$O}
\label{sec:inheritance}
Comets are thought to be the most pristine objects of the cold ice-bearing regions in the solar nebula.
Observations of cometary comae indicate that the $\hdo$ ratio ranges from $3\times10^{-4}$ to $11\times10^{-4}$ depending on the source 
\citep[with no clear difference in the $\hdo$ ratio between Oort Cloud comets and Jupiter-family comets; e.g.,][]{mumma11,altwegg15}.
Based on these measurements, there are long-standing arguments on the origin of cometary water \citep[e.g.,][]{geiss81,aikawa99}.
The suggested possibilities range from prestellar inheritance to in-situ formation in protoplanetary disks as two extremes 
\citep[e.g.,][for recent modeling work]{furuya13,albertsson14,cleeves14}.
The main difficulty in distinguishing between these two cases comes from the fact that deuterium fractionation of molecules driven by isotopic exchange reactions,
such as $\reacteq{H_3^+}{HD}{H_2D^+}{H_2}$, can occur efficiently both in the prestellar stages and in the cold midplane of protoplanetary disks \citep[e.g.,][]{aikawa99}.

\subsubsection{Inheritance versus in-situ formation in disks}
Our results (in part) support the prestellar inheritance scenario; 
interstellar water ice is largely delivered to forming disks without significant alteration.
When averaged over the whole disk, our forming disk has an H$_2$O abundance and $\hdo$ ratio similar to the precollapse values.
On the other hand, due to the selective loss of HDO ice with respect to H$_2$O ice, the local vertically averaged $\hdo$ ice ratio can be more divergent 
depending on time, distance from the central star, and the rotational rate of the prestellar core.
Our fiducial model reproduces the variation in the $\hdo$ ratio observed in
comets via the combination of ice formation in the prestellar stages and the selective loss of HDO en route into the disk (see Figure \ref{fig:column}).

However, even if this is the case, chemical processing in forming disks themselves can eliminate the prestellar inheritance.
At high gas temperatures, isotopic exchange between HDO and H$_2$ in the gas phase, $\reacteq{HDO}{H_2}{H_2O}{HD}$, is thermally activated, and 
in equilibrium at $T_{\rm gas} \gtrsim 500$ K, the $\hdo$ ratio is close to the HD/H$_2$ ratio, i.e., $2\times \dhr{elem}$ \citep{richet77}. 
The rate coefficient of the forward reaction at $>$300 K was measured  by \citet{lecluse94}\footnote{
More precisely, they measured the rate coefficient for the D$_2$O-H$_2$ system, and that for the HDO-H$_2$ system was estimated from the assumption of statistical branching ratios.}, 
and the timescale of the forward reaction is given by $\tau \approx 10\exp(5170\, {\rm K}/T_{\rm gas})$ (10$^{13}\, {\rm cm^{-3}}$/$n_{\rm H_2}$) yr, 
e.g., at $T_{\rm gas}$ = 500 K and $n_{\rm H_2}$ = 10$^{13}$ cm$^{-3}$, $\tau \approx 0.5$ Myr.
Thus the impact of this reaction is negligible on the timescale of star and disk formation, 
except for regions with both very high density and very high temperature,  i.e., the inner regions of forming disks.
Such reprocessed water could be mixed with unprocessed interstellar water via radial spreading and accretion \citep{drouart99,yang13}.
The resulting radial profile of the $\hdo$ ratio depends on how much material is transported from the very hot, dense inner regions to the outer cold regions.
The presence of crystalline silicates in comets in our solar system \citep[e.g.,][]{zolensky06} could imply that 
water was also efficiently thermally processed in the solar nebula \citep[but see e.g.,][]{vinkovic09,tazaki15}.

The work by \citet{cleeves14} puts limits on the efficiency of the thermal processing in the solar nebula.
They show that if water deuteration is completely reset by processing (i.e., $\hdo = 2\times \dhr{elem}$, but volatile oxygen is mostly locked up in H$_2$O and CO), 
ionization and ice chemistry in Class II disks cannot produce enough HDO to explain the measured level of water deuteration in the solar system.
The main limiting factor for the HDO production is the availability of volatile oxygen \citep[see also][]{willacy09,furuya13}.
The claim is further strengthened if the cosmic rays are modulated by the stellar wind \citep[][]{cleeves13,cleeves14}.
Therefore, at least some fraction of disk water would avoid the high-temperature processing in the solar nebula.
In summary, it is now possible to explain the $\hdo$ measurements in comets quantitatively by the prestellar inheritance scenario being supported by astrophysical and chemical models.

On the other hand, it is also possible to explain the $\hdo$ measurements in comets via Class II disk chemistry with turbulent mixing, regardless of the initial disk compositions.
\citet{furuya13} and \citet{albertsson14} independently claim that abundant water ice with cometary $\hdo$ ratios can be formed 
in their Class II disk chemical models with turbulent mixing, regardless of the initial $\hdo$ ratio.
Turbulence can bring ice-coated dust grains from the disk midplane to the disk surface where stellar UV and X-ray radiation destroy ices.
Similarly, turbulence can bring atomic oxygen from the disk surface to the midplane, where volatile oxygen is largely locked up in the ice.
The cycle of destruction and reformation resets the chemical composition, and then the steady state composition in the turbulent disk model 
does not depend on the initial composition, but does depend on the thermal structure of disks.
The models predict that the $\hdo$ ice ratio increases with radius reflecting the thermal structure of disks.
As a cautionary note, it is unclear if the turbulent disk chemistry can reach steady-state in reality.
The models of \citet{furuya13} and \citet{albertsson14} did not consider the effect of gas accretion onto the central star and dust evolution, 
which can lead to non-equilibrium chemical compositions.
Furthermore, recent observations suggest that the turbulence in the surface layers of disks is not strong (non-thermal motion is less than 3 \% of the local sound speed), 
meaning that turbulent mixing may also be weaker than commonly assumed in models \citep{flaherty15}.

\subsubsection{The ratio of $\ddo$ to $\hdo$}
The main question here is again how we can distinguish observationally between inheritance and in-situ formation of water in a protoplanetary disk.
To complicate matters further, our models show that even if interstellar water ice is destroyed by stellar UV and reformed prior to the disk entry, 
the $\hdo$ ratio in reformed water ice is similar to the original value (Figure \ref{fig:uv_ab_case7}).
One way could be to constrain the disk radial profile of the $\hdo$ ratio;
the in-situ formation scenario predicts that the $\hdo$ ratio is an increasing function of distance from central stars, 
while the inheritance scenario predicts that the $\hdo$ ratio does not always increase with radius depending on parameters.
Another way is to look for probes of prestellar inheritance other than the $\hdo$ ratio.

A possible alternative is the ratio of $\ddo$ to $\hdo$ (hereafter $f_{\rm D2}/f_{\rm D1}$ ratio).
In contrast to the $\hdo$ ratio, the $f_{\rm D2}/f_{\rm D1}$ ratio is significantly different between the original and reformed water ices 
(10 versus $\sim$0.1; see the bottom panel of Figure \ref{fig:uv_ab_case7}).
The water observations toward low-mass protostellar cores have suggested that the water ice formed in the prestellar stages 
may be characterized by a high $f_{\rm D2}/f_{\rm D1}$ ratio of $>$1.
\citet{coutens14} quantified the $f_{\rm D2}/f_{\rm D1}$ ratio to be $\sim$7 in the hot inner regions ($>$100 K) around the Class 0 protostar 
NGC 1333-IRAS 2A, where water ice has sublimated.
Quasi-steady state chemistry of water ice formation on grain surfaces leads to $f_{\rm D2}/f_{\rm D1} \le 0.25$ \citep[cf.][]{rodgers02}.
\citet{furuya16b} proposed that the anomalously high $f_{\rm D2}/f_{\rm D1}$ ratio is a natural consequence of the chemical evolution during low-mass star formation:
(i) the majority ($\gtrsim$90 \%) of volatile oxygen is locked up in water ice and other O-bearing molecules without significant deuterium fractionation and 
(ii) at later times, water ice formation continues with reduced efficiency but with enhanced deuterium fractionation 
(the probability of deuteration with respect to hydrogenation is enhanced by a factor of $\gtrsim$100).
The enhancement of deuterium fractionation can be triggered by a drop in the ortho-to-para nuclear spin ratio of H$_2$, CO freeze-out, 
and the attenuation of the interstellar UV field.
When either of the two conditions is not satisfied, the $f_{\rm D2}/f_{\rm D1}$ ice ratio is close to 0.25 or smaller.
In the warm infalling protostellar envelope, it is difficult to satisfy both conditions, especially condition (ii).
Therefore, the reformed water ice in the warm envelope has a lower $f_{\rm D2}/f_{\rm D1}$ ratio than the original water ice.

The chemical processing in disks discussed above leads to the $f_{\rm D2}/f_{\rm D1}$ ratio of around unity or smaller.
The thermally induced exchange reactions $\reacteq{XDO}{H_2}{XHO}{HD}$, where X is H or D, in the inner regions of forming disks would 
lead to $f_{\rm D2}/f_{\rm D1}$ of unity in equilibrium at high temperatures ($>$500 K).
We checked the Class II turbulent disk chemical model of \citet{furuya13} and found that the $f_{\rm D2}/f_{\rm D1}$ ratio is around unity or smaller 
in regions where abundant water ice is present reproducing  the cometary $\hdo$ ratios (see Appendix \ref{appendix:mixing}).
Taken together, the $f_{\rm D2}/f_{\rm D1}$ ratio better probes the prestellar inheritance of H$_2$O than the $\hdo$ ratio.
The radial profiles of the $f_{\rm D2}/f_{\rm D1}$ ice ratio in our models are shown in the bottom panels of Figures \ref{fig:column} and \ref{fig:column_c3}.
The $f_{\rm D2}/f_{\rm D1}$ ice ratio is much larger than unity except for the regions where the H$_2$O ice abundance is low, 
i.e., the $f_{\rm D2}/f_{\rm D1}$ ratio is a measure of how much H$_2$O was lost during the disk formation.
The comparisons between the observationally derived $f_{\rm D2}/f_{\rm D1}$ ratio in the bulk ice in clouds/cores and that in disks/comets would provide the strongest constraints  
on the prestellar inheritance of H$_2$O.
There is no detection of D$_2$O in comets yet in the literature.
In addition, whether $f_{\rm D2}/f_{\rm D1} > 1$ is common in low-mass star forming regions is not fully confirmed observationally.
More D$_2$O observations toward protostars and comets with detections of H$_2$O and HDO are highly desired.

\subsection{Water deuteration versus organics deuteration}
\label{sec:organics}
Observations toward the inner hot regions ($>$100 K) around Class 0 low-mass protostars, where ices have sublimated, have found 
that formaldehyde and methanol show higher levels of deuterium fractionation 
than water typically by a factor of 10 or more \citep[e.g.,][]{parise06,coutens13,persson14}.
The inferred gas-phase fractionation ratios are thought to reflect those in the ice through thermal desorption.
The idea is supported by 1D collapsing core models with detailed gas-ice chemical models \citep[e.g.,][]{aikawa12,taquet14}.
We confirmed that our 2D model at early times ($t = 0.3t_{\rm acc}$), which considers the effect of stellar UV radiation unlike the 1D models, 
also supports this idea, although the size of the hot regions in our model is more compact ($\sim$10 au) 
than the size of the emission estimated from the observations \citep[several tens of au;][]{persson13,maury14,taquet15}.
Thus it is likely that the icy organics have higher deuteration than water ice in the ISM.
This trend likely reflects the different epoch of their formation, i.e., H$_2$O ice is formed earlier than HDO,
formaldehyde and methanol ices during star formation.
Recent astrochemical models have successfully reproduced this trend \citep[][]{taquet14,furuya16b}.

The main question here is whether or not the higher deuteration in organics is transferred to circumstellar disks.
Hereafter we use the D/H ratio of molecules instead of the abundance ratio between the regular and deuterated forms; the former is often used in the solar system community.
For molecule H$_n$X, $\dhr{H_nX} = n^{-1} \times ({\rm H}_{n-1}{\rm DX/H}_{n}{\rm X)}$.
Then $\dhr{H_2O} = 0.5 \times (\hdo)$.
Observations of cometary comae have quantified $\dhr{H_2O}$ at $\sim (1-6)\times10^{-4}$ \citep[e.g.,][]{altwegg15}.
The detection of other deuterated molecules in comae is not reported except for DCN in comet Hale-Bopp.
The measured D/H ratio of HCN is $2\times10^{-3}$, which is higher than that of water in the same comet by a factor of $\sim$7 \citep{meier98a,meier98b}.
The upper limits of the D/H ratios of methanol for CH$_2$DOH and for CH$_3$OD in comet Hale-Bopp are $8\times10^{-3}$ and 0.025, respectively \citep{crovisier04}.
Meanwhile, like comets, chondritic meteorites are thought to be pristine probes of the solar nebula.
In meteorites, both hydrated minerals (i.e., water) and organic materials contribute to the bulk D/H ratio.
\citet{alexander12} found that the bulk D/H ratio increases with the bulk C/H ratio in a sample of meteorites, 
which indicates organics are more enriched in deuterium than water \citep[e.g.,][for reviews]{robert06,ceccarelli14}.
In summary, these measurements indicate that organics in pristine objects of the solar system are more enriched in deuterium than water, 
although the available data are limited for comets. 

\begin{figure}
\resizebox{\hsize}{!}{\includegraphics{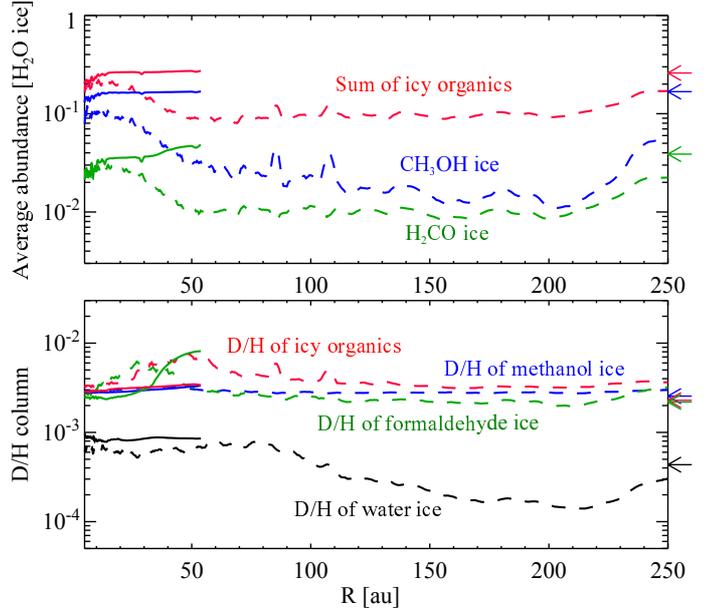}}
\caption{Molecular abundances and D/H ratios in the disk at  $t=0.5t_{\rm acc}$ (solid lines) and $t=t_{\rm acc}$ (dashed lines) 
as functions of radius in our fiducial model.
Top: vertically averaged abundances of H$_2$CO ice, CH$_3$OH ice, and the sum of icy organics with respect to H$_2$O ice.
Bottom: the D/H column density ratios of formaldehyde ice, methanol ice, icy organics, and water ice.
Arrows on the right-hand margin indicate the values at the onset of collapse.
}
\label{fig:column_organics}
\end{figure}

The top panel of Figure \ref{fig:column_organics} shows the vertically averaged abundances with respect to H$_2$O ice of H$_2$CO ice, CH$_3$OH ice, and 
the sum of icy organics as functions of radius at $t=0.5t_{\rm acc}$ and $t=t_{\rm acc}$ in the fiducial model.
The total abundance of icy organics was calculated by summing the abundances of icy molecules with at least one C-H bond.
The lower panel shows their D/H column density ratios.
The D/H ratio of methanol ice is calculated by $0.25 \times (x_{\rm CH_2DOH_{ice}} + x_{\rm CH_3OD_{ice}})/x_{\rm CH_3OH_{ice}}$, 
because our chemical network does not fully distinguish the various possible isomers of radicals, such as CH$_2$DO and CH$_2$OD, which are relevant for methanol.
The CH$_2$DOH/CH$_3$OD ice ratio in our model is close to the statistical value of 3 due to the incomplete distinction of the isomers.
Also, our network does not include the substitution and abstraction reactions of formaldehyde and methanol.
These reactions can enhance the level of deuterium fractionation in formaldehyde and methanol, and can affect the abundance ratio of methanol isomers \citep{watanabe08}.
In order to check the impact of our simplified treatment of icy methanol chemistry, we reran the molecular cloud formation model and the prestellar core model
(Section \ref{sec:init}) with a modified network.
The modification includes (i) the distinction of  the possible isomers, and (ii) the addition of the substitution and abstraction reactions following the method of \citet{taquet12}.
The modified network gives a higher CH$_2$DOH/CH$_3$OD ice ratio (5 versus 3) and a higher D/H ratio of methanol ($1\times10^{-2}$ versus $3\times10^{-3}$) 
at the onset of collapse compared to our fiducial network.
The D/H ratios of icy organics are calculated from the total number of D atoms in all icy organics divided by the total number of H atoms in all icy organics as in \citet{cleeves16}.

At $t=0.5t_{\rm acc}$, the CH$_3$OH ice abundance and the total abundance of icy organics are similar to their precollapse values.
At $t=t_{\rm acc}$, the CH$_3$OH ice abundance ranges from 1 to 10 \% of water ice, which is lower than its precollapse value \citep[see also][]{drozdovskaya14}. 
CH$_3$OH ice is selectively lost with respect to H$_2$O ice, since CH$_3$OH is mostly present in the upper ice layers 
in contrast to H$_2$O which is mostly present in the lower layers (Figure \ref{fig:init_chem}).
This is true for H$_2$CO ice as well.
The D/H column density ratios of icy organics in the disk are similar to or slightly higher than their precollapse values, regardless of time and radius, 
and are higher than the D/H ratio of water ice by a factor of 3 or more. 
As both regular and deuterated forms of methanol ice are mostly present in the upper ice layers, 
selective loss of deuterated methanol relative to its regular isotopologue does not occur unlike water, even if materials are subject to stellar UV radiation.
The differences between the D/H ratios of the icy organics and water ice tend to remain or even become enhanced during the collapse and en route into the disk. 

\section{Summary}
\label{sec:summary}
We have investigated the delivery of regular and deuterated water from prestellar cores to circumstellar disks.
We used the axisymmetric 2D model originally developed by \citet{visser09b,visser11} to follow the physical evolution of collapsing cores.
In the physical simulations, streamlines of fluid parcels were tracked, and gas-ice astrochemical simulations were performed along 
the streamlines.
Our findings are summarized as follows.

\begin{enumerate}
{\item
There are three cases for water ice chemistry in the infalling envelope, depending on the degree of processing by stellar UV photons that the infalling fluid parcels experience:
(i) the prestellar ice is preserved, and both the H$_2$O ice abundance and the $\hdo$ ice ratio are unaltered during the collapse,
(ii) only the upper layers of the prestellar ice are lost and then the H$_2$O ice abundance is unaltered, while the $\hdo$ ice ratio is lowered (cf. Figure \ref{fig:init_chem}), and
(iii) the entire ice mantle is largely lost (and partly reformed).
In the last case,  the $\hdo$ ratio in reformed water ice is similar to the original value, because the photofragments are enriched in deuterium.
}


{\item
Our models suggest that interstellar water ice is largely delivered to forming disks without significant alteration.
When averaged over the whole disk, the H$_2$O ice abundance and $\hdo$ ice ratio are similar to the precollapse values to within a factor of $\sim$2, 
regardless of time in our fiducial model. 
On the other hand, the local vertically averaged H$_2$O ice abundance and $\hdo$ ice ratio shows larger differences depending on time 
and distance from the central star, reflecting the variation of the composition of the accreting material.
}

{\item 
How the chemical differentiation in the protostellar envelope is reflected in the forming disk depends on the solid-body rotation rate of the core in our models, 
which in turn determines the UV fluence and temperature variations that the infalling fluid parcels experience.
As the outflow direction and the rotational axis are aligned, the more significantly processed material by the stellar UV tends to have a lower specific angular momentum.
In the model with lower core rotation rate ($\Omega=10^{-14}$ s$^{-1}$;  in our fiducial model, 10$^{-13}$ s$^{-1}$),  
the envelope material which was processed via the stellar UV radiation most likely eventually accreted onto the small spatial scales (i.e., star) 
or enters into the outflow cavity and does not contribute to the disk material.
}

{\item
There are long-standing arguments on the origin of cometary water based on the $\hdo$ ratio.
The suggested possibilities range from prestellar inheritance to in-situ formation in protoplanetary disks.
We propose that the ratio of $\ddo$ to $\hdo$ is a better probe to distinguish the two cases than solely the $\hdo$ ratio (see Section \ref{sec:inheritance}).
The comparison between the observationally derived ratio in the bulk ice in clouds/cores and that in disks/comets would provide the strongest constraints 
on the origin of cometary water.
Future D$_2$O observations with H$_2$O and HDO toward hot gas ($>$100 K) around protostars and toward cometary comae are crucial.
}

{\item
Icy organics are more enriched in deuterium than water ice in the forming disks in our models.
The trend is inherited from the prestellar stages.
}
\end{enumerate}

 
\begin{acknowledgements}
We thank the anonymous referee for useful comments that helped to improve this paper.
Astrochemistry in Leiden is supported by the Netherlands Research School for Astronomy (NOVA), by a Royal Netherlands Academy of Arts
and Sciences (KNAW) professor prize, and by the European Union A-ERC grant 291141 CHEMPLAN.
K.F. acknowledges support by the Research Fellowship from the Japan Society for the Promotion of Science (JSPS).
M.N.D. acknowledges support by a Huygens fellowship from Leiden University.
C.W. acknowledges financial support from the Netherlands Organisation for Scientific Research (NWO, program number 639.041.335).
D.H. is funded by Deutsche Forschungsgemeinschaft Schwerpunktprogramm (DFG SPP 1385) The First 10 Million Years of the Solar System – a Planetary Materials Approach.
Numerical computations were, in part, carried out on PC cluster at Center for Computational Astrophysics, National Astronomical Observatory of Japan.
\end{acknowledgements}

\begin{appendix}
\section{(2+$n$)-phase gas-ice astrochemical model}
\label{appendix:mph}
In order to simulate molecular evolution in the dense ISM appropriately, one must consider, at least, 
two phases of chemical species, gas and ice \citep[the so-called two-phase model;][]{hasegawa92}.
In the original two-phase model, the whole ice mantle is assumed to be chemically active, while laboratory experiments have shown that
two-body association reactions to form icy molecules preferentially occur in the surface layers rather than the whole ice mantle at low temperatures \citep[e.g.,][]{watanabe07,ioppolo10}.
In order to consider the different chemical activity, \citet{hasegawa93b} proposed the so-called three-phase model, in which three phases, gas, icy surface, and the bulk ice, are considered.
The basic assumption of the three-phase model is that the bulk ice mantle has uniform chemical composition, while ice composition in the ISM is most likely inhomogenous.
In order to take into account a depth-dependent ice structure, we introduce a (2+$n$)-phase model, where $n \ge 2$, as a natural extension of the three-phase, i.e., (2+1)-phase, model.

Let us start from describing the three-phase model originally proposed by \citet{hasegawa93b}.
The three-phase model consists of three sets of ordinary differential equations (ODEs): 
\begin{align}
\req{\dot{n}}{i}{g} &= \req{P}{i}{g} - \req{L}{i}{g}, \\ \label{eq:3ph1}
\req{\dot{n}}{i}{s} &= \req{P}{i}{s} - \req{L}{i}{s} - T_i, \\
\req{\dot{n}}{i}{m} &= \req{P}{i}{m} - \req{L}{i}{m} + T_i, \label{eq:3ph3}
\end{align}
where $\req{n}{i}{g}$, $\req{n}{i}{s}$, and $\req{n}{i}{m}$ are the number densities of species $i$ per unit gas volume 
in the gas phase, on the surface of the icy dust grain, and in the ice mantle, respectively.
The dots represent the derivatives with respect to time.
$P_i$ and $L_i$ are the production and loss rates of species $i$ by chemical processes, respectively.
In the three-phase model, molecules on the surface layers become part of the ice mantle with the growth of ice,
while part of the ice mantle become the surface layers with the loss of ice.
We denoted the surface-mantle transition rate of species $i$ as $T_i$. 
The total rate of the surface-mantle transition, $\sum_{i}T_i$, 
should satisfy $0 \le |\sum_{i}T_i| \le |R^{(s)}|$, where $R^{(s)} = \sum_{i}[\req{P}{i}{s} - \req{L}{i}{s}]$.
$R^{(s)} \geq 0$ and $R^{(s)} < 0$ correspond to the net growth and net loss of ice, respectively.
One may write
\begin{align}
  \sum_{i}T_i =  R^{(s)} \times \begin{cases}
    f_{\rm grow} \, , \,\,\,\,\, R^{(s)} \geq 0, \\
    f_{\rm loss} \, , \,\,\,\,\, R^{(s)} < 0,
  \end{cases}
\end{align}
where $0 \le f_{\rm grow} \le 1$ and $0 \le f_{\rm loss} \le 1$.
We denote $\sum_{k}\req{n}{k}{x}$ as $\sumreq{n}{x}$.
It is often assumed that chemical processes occur only in the uppermost monolayer.
In this case $f_{\rm grow}$ is given as the coverage of a surface, i.e., $\sumreq{n}{s}/(N_{\rm site}n_{\rm gr}$), 
where $N_{\rm site}$ is the number of binding sites per monolayer and $n_{\rm gr}$ is the number density of dust grains per unit gas volume \citep{hasegawa93b}.
$\sumreq{n}{s}/n_{\rm gr}$ represents the average number of icy molecules on the surface of a dust grain.
On the other hand, we consider the uppermost $N_{\rm actlay}$ layers to be chemically active ($N_{\rm actlay} = 4$ MLs in this paper), and
$f_{\rm grow}$ is given by [$\sumreq{n}{s}/(N_{\rm site}n_{\rm gr}) - (N_{\rm actlay}-1)$] with the lower limit of zero.
Then the surface-mantle transition occurs after the formation of ($N_{\rm actlay}-1$) monolayers of ice in our model.  
$f_{\rm loss}$ is $\sumreq{n}{m}$/min$(\sumreq{n}{s}, N_{\rm site}n_{\rm gr})$ with the upper limit of unity \citep{garrod11}.
To determine each $T_i$, it is assumed that the rate is proportional to the fractional composition:
\begin{align}
  T_i =  R^{(s)} \times \begin{cases}
    f_{\rm grow}\req{n}{i}{s}/\sumreq{n}{s} \, , \,\,\,\,\, R^{(s)} \geq 0, \\
    f_{\rm loss}\req{n}{i}{m}/\sumreq{n}{m} \, , \,\,\,\,\, R^{(s)} < 0.
  \end{cases}
\end{align}

In Equation (\ref{eq:3ph3}) we did not consider diffusion between the surface and the bulk ice mantle,
which would be important for the segregation of CO$_2$ ice as suggested from the ice observations in protostellar envelopes, 
although the mechanism and the efficiency in the ISM ice remain uncertain \citep{oberg09,fayolle11}.

The basic assumption of the three-phase model is that the bulk ice mantle has uniform chemical composition.
In order to take into account a depth-dependent ice structure, we divide the bulk ice mantle into distinct $n$ phases, where $n \ge 2$.
One may refer to the model as a (2+$n$)-phase model analogously to the (2+1)-phase (three-phase) model.
The (2+$n$)-phase model consists of 2+$n$ sets of ODEs:
\begin{align}
\req{\dot{n}}{i}{g} &= \req{P}{i}{g} - \req{L}{i}{g}, \\ 
\req{\dot{n}}{i}{s} &= \req{P}{i}{s} - \req{L}{i}{s} - \sum_{k}\req{T}{i}{m_k}, \\
\req{\dot{n}}{i}{m_j} &= \req{P}{i}{m_j} - \req{L}{i}{m_j} + \req{T}{i}{m_j} \,\,(j=1,2,3..,n), \label{eq:mph3}
\end{align}
where $\req{n}{i}{m_j}$ is the number density of species $i$ per unit gas volume in the $j$-th mantle phase ($m_j$).
We refer to the lowermost mantle phase as the 1st mantle ($m_1$), 
while the uppermost mantle phase as the $n$-th mantle ($m_n$).
Each mantle phase consists of $N^{(m_j)}_{\rm maxlay}$ ice monolayers at a maximum.
It is assumed that when the net growth of ice occurs, the mantle phases are filled in order of $m_1$ $\rightarrow$ $m_2$ $\rightarrow$... $\rightarrow$ $m_n$,
while when the net loss occurs, the mantle phases are emptied in reverse.
The surface-mantle transition rate of species $i$ in the $j$-th mantle is given as follows:
\begin{align}
  \req{T}{i}{m_j} =  R^{(s)} \times \begin{cases}
    \alpha^{(m_j)}_{\rm grow} f_{\rm grow}\req{n}{i}{s}/\sumreq{n}{s} \, , \,\,\, R^{(s)} \geq 0, \\
    \alpha^{(m_j)}_{\rm loss} f_{\rm loss}\req{n}{i}{m_j}/\sumreq{n}{m_j} \, , \,\,\, R^{(s)} < 0,
  \end{cases}
\end{align}
where $f_{\rm loss}$ was redefined by $\sum_{j}\sumreq{n}{m_j}$/min$(\sumreq{n}{s}, N_{\rm site}n_{\rm gr})$ with the upper limit of unity.
We introduced additional factors, $\alpha^{(m_j)}_{\rm grow}$ and $\alpha^{(m_j)}_{\rm loss}$, which describe "activity" of the $j$-th mantle for the surface-mantle transition, and 
satisfy $0 \le \alpha^{(m_j)}_{\rm grow} \le 1$, $0 \le \alpha^{(m_j)}_{\rm loss} \le 1$, and $\sum_{j}\alpha^{(m_j)}_{\rm grow} = \sum_{j}\alpha^{(m_j)}_{\rm loss} = 1$.
Then, $\sum_{j}\sum_{i}\req{T}{i}{m_j}$ is identical to $\sum_{i}T_i$ in the three-phase model;
if one is interested only in the formation of ice and assumes that bulk ice mantle is chemically inert, 
the three-phase model and the (2+$n$)-phase model with arbitrary $n$ give identical whole ice compositions.
The functional forms of $\alpha^{(m_j)}_{\rm grow}$ and $\alpha^{(m_j)}_{\rm loss}$ are given as follows:
\begin{align}
\alpha^{(m_j)}_{\rm grow} &= 1 - {\rm min}[1, \,\, (N^{(m_{j-1})}_{\rm maxlay} - \sumreq{n}{m_{j-1}}/(N_{\rm site}n_{\rm gr}))], \\
\alpha^{(m_j)}_{\rm loss} &= 1 - {\rm min}[1, \,\, \sum_{k}^{k > j}\sumreq{n}{m_k}/(N_{\rm site}n_{\rm gr})].
\end{align}

Finally, we note that there is a way to treat ice structure in layer-by-layer manner even within the framework of the two-phase model 
by keeping track of each layer as it forms \citep{taquet12}.
However, it was found that the method is computationally expensive when simulating ice sublimation \citep{taquet14}, 
loosing the major advantage of the rate equation method over more detailed methods like the macroscopic Monte-Carlo approach \citep{vasyunin13}.
As has been demonstrated in this work, our (2+$n$)-phase model can be used with a complicated chemical network and a time-dependent physical model, 
which covers wide ranges of physical conditions, although the resolution of the depth-dependent ice structure in our model is limited.
In principle, it is possible to treat each monolayer as a distinct phase in the (2+$n$)-phase model by setting to $N^{(m_j)}_{\rm maxlay} = 1$.
It requires, however, significant computational effort because of the increasing number of ODEs to be solved.
Another advantage of our (2+$n$)-phase model is being able to consider chemical processes in the distinct mantle phases as it has been done in Section \ref{sec:bulkice},
which is beyond the framework of the two- and three-phase models.

\section{Additional figures from the model with a lower solid-body rotation rate of the core}
\label{appendix:rotation}
\begin{figure*}
\sidecaption
\resizebox{\hsize}{9cm}{\includegraphics{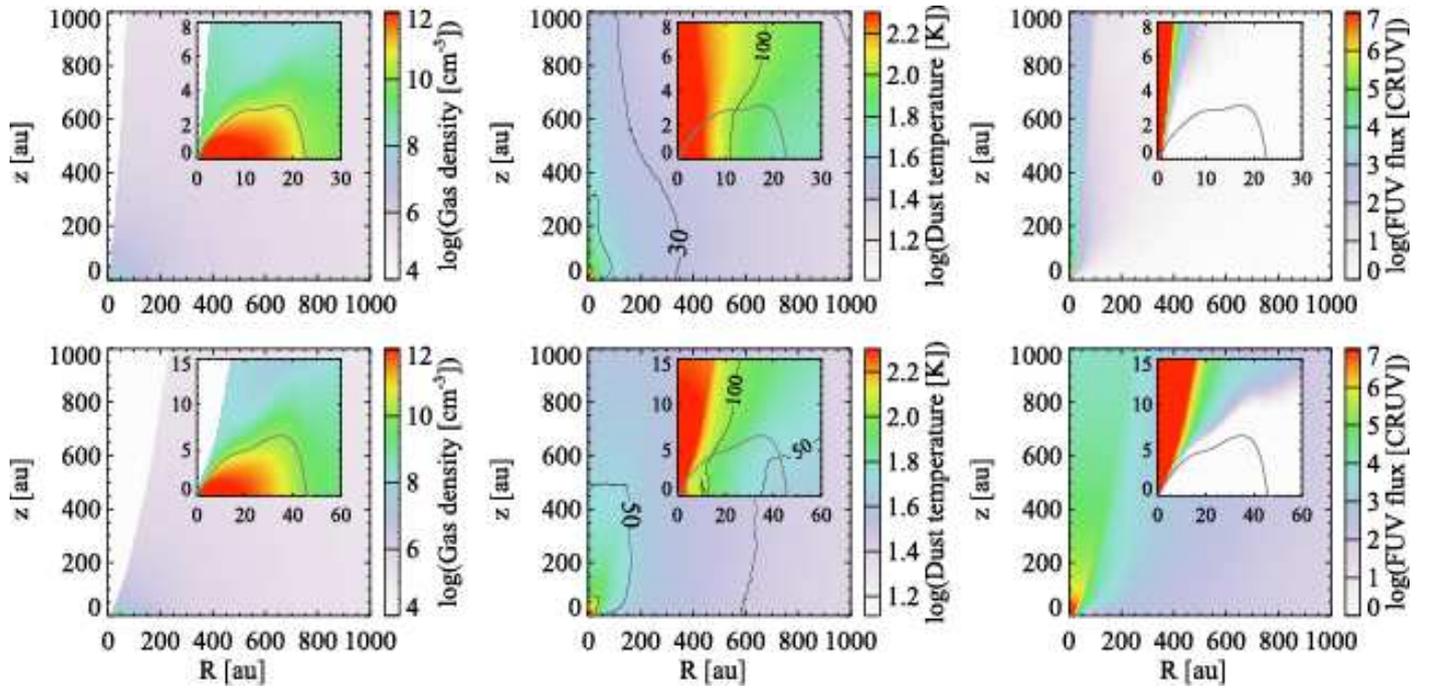}}
\caption{Spatial distributions of the gas density (left), dust temperature (middle), and stellar UV radiation field 
at $t = 0.5 t_{\rm acc}$ (top) and $0.9t_{\rm acc}$ (bottom) in model the model with low core rotation rate ($\Omega = 10^{-14}$ s$^{-1}$).
Other details are the same as Figure \ref{fig:phys}.
}
\label{fig:phys_c3}
\end{figure*}

\begin{figure*}
\resizebox{12cm}{!}{\includegraphics{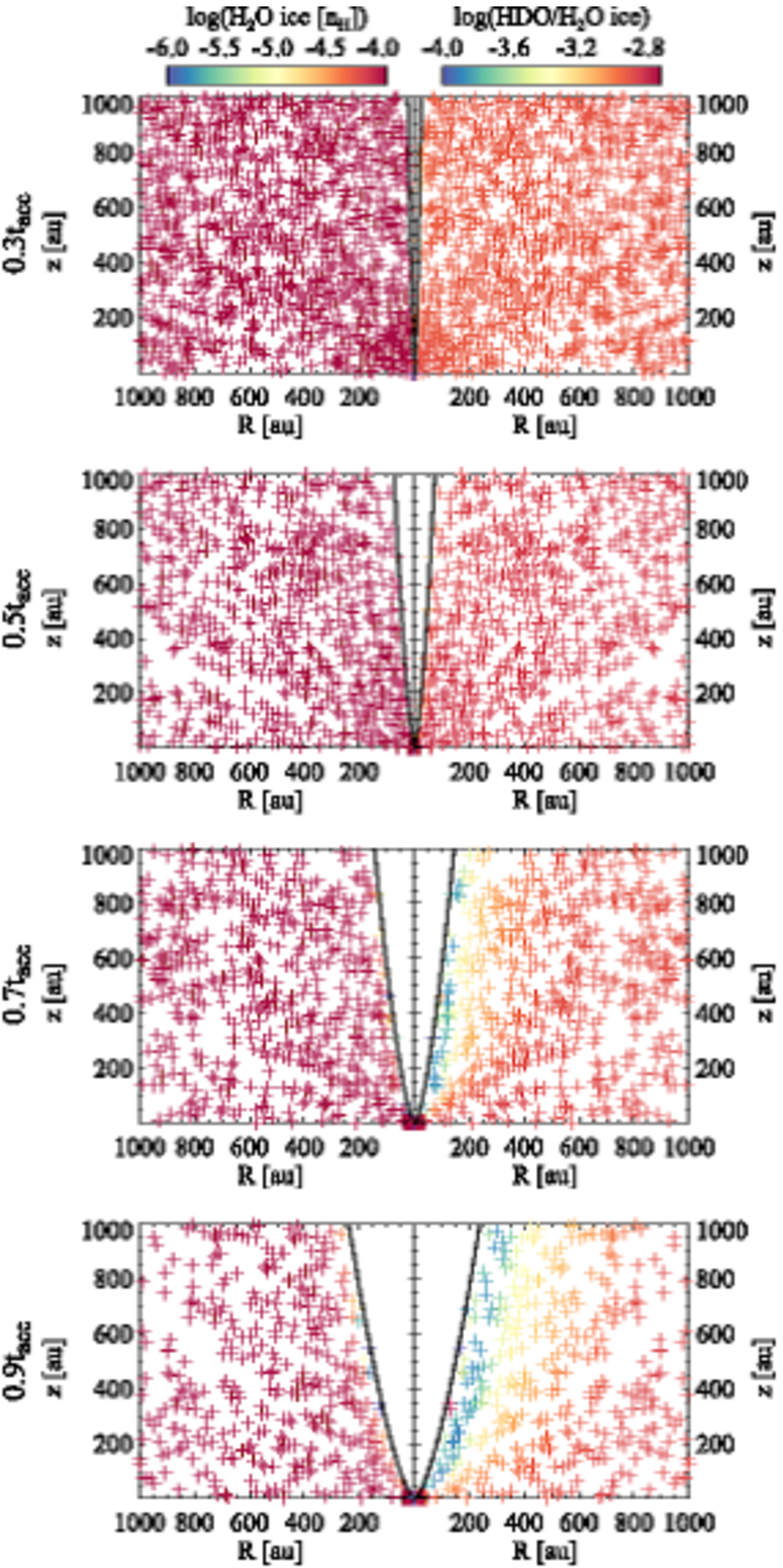}}
\caption{Spatial distributions of fluid parcels on a 1000 au scale at 0.3$t_{\rm acc}$, 0.5$t_{\rm acc}$, 0.7$t_{\rm acc}$, and
0.9$t_{\rm acc}$ (from top to bottom) in the model with a low core rotation rate ($\Omega = 10^{-14}$ s$^{-1}$). 
Left: H$_2$O ice abundance with a logarithmic scale.
Right: icy $\hdo$ ratio with a logarithmic scale.
The black solid lines at each panel depict the disk surface and the outflow cavity wall.
See Figure \ref{fig:chem_c3_zoom} for the zoom-in view.
}
\label{fig:chem_c3}
\end{figure*}

\begin{figure*}
\resizebox{12cm}{!}{\includegraphics{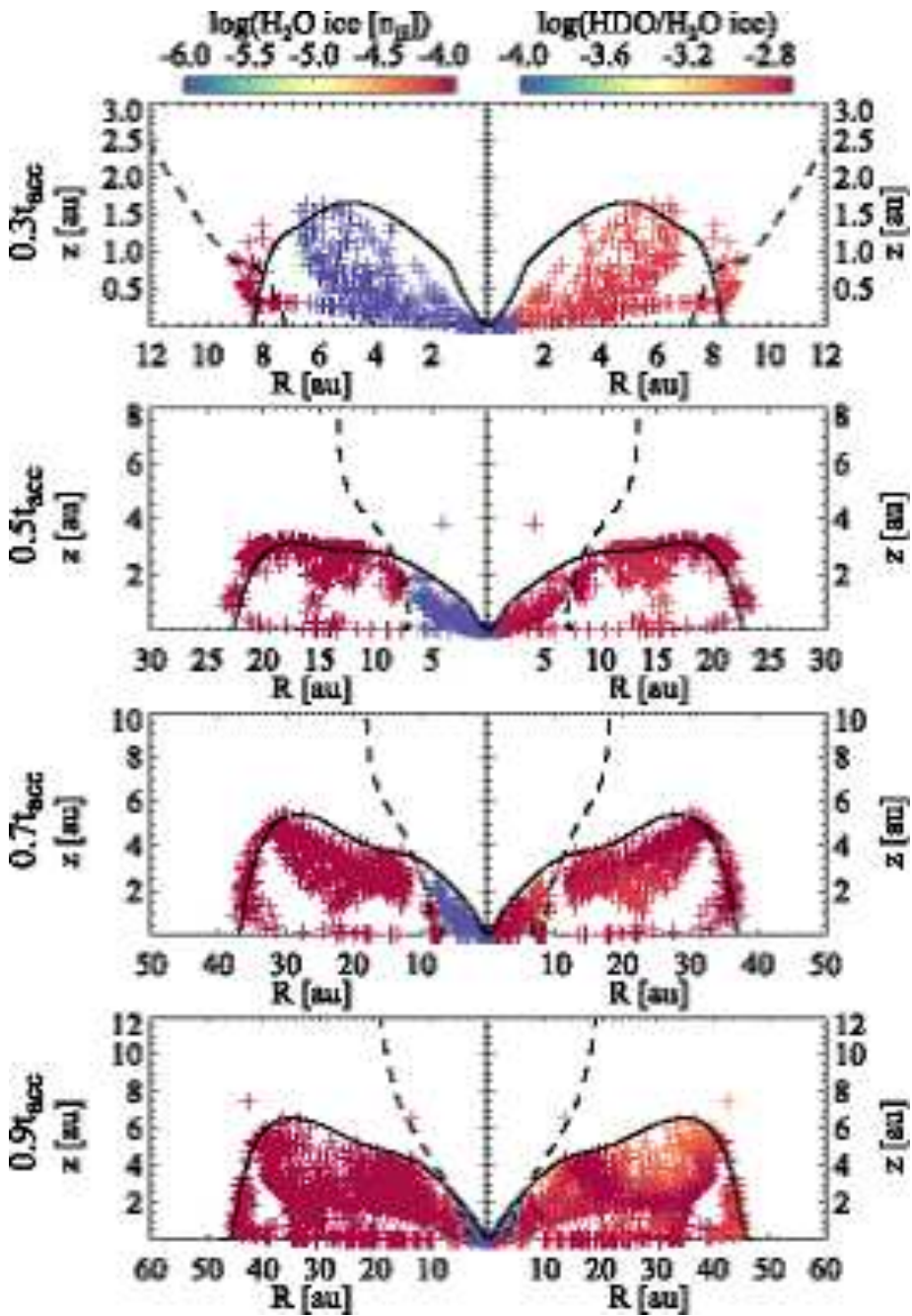}}
\caption{Zoom-in view of Figure \ref{fig:chem_c3}. Note the different spatial scale among the panels.
Water snow lines are depicted by dashed lines.
The outflow cavity walls are not shown in this figure.
}
\label{fig:chem_c3_zoom}
\end{figure*}

\begin{figure*}
\resizebox{12cm}{!}{\includegraphics{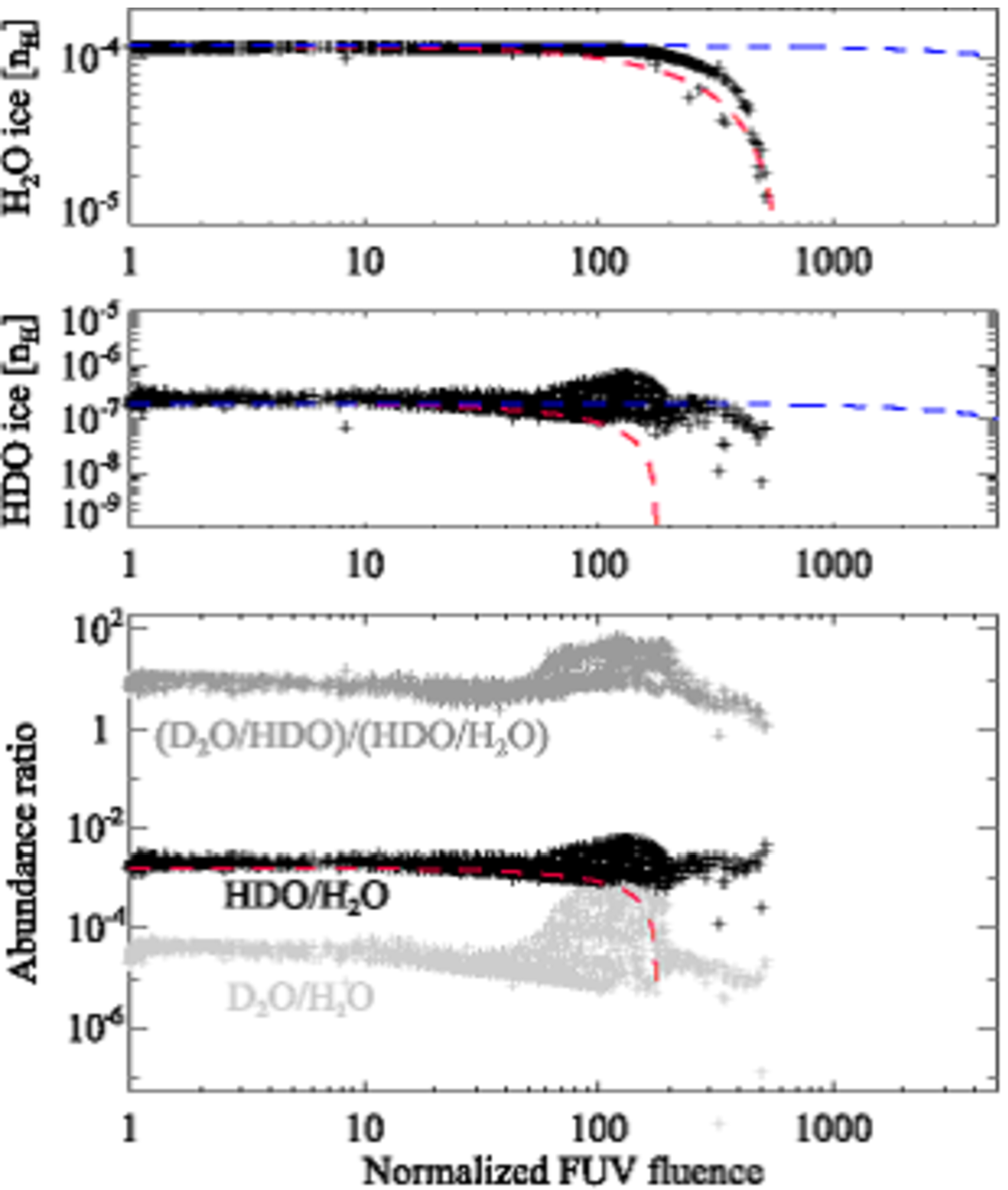}}
\caption{H$_2$O ice (top)  and HDO ice (middle) abundances in the fluid parcels that are located in the disk at $t=t_{\rm acc}$ 
as functions of the normalized UV fluence in the model with a low core rotation rate ($\Omega = 10^{-14}$ s$^{-1}$).
The bottom panel shows $\hdo$ ice ratio (black), D$_2$O/H$_2$O ice ratio (light gray), and the ratio of $\ddo$ to $\hdo$ ice ratio (gray).
Only the fluid parcels in which the H$_2$O ice abundance is larger than 10$^{-5}$ are plotted.
The blue and red dashed lines depict the expected H$_2$O (or HDO) ice abundance when only photodesorption is 
an allowed chemical process and when only photodissociation/photodesorption are included, respectively.
}
\label{fig:uv_ab_case3}
\end{figure*}

Here we present additional figures from the model with a solid-body rotation rate of the core of $\Omega=10^{-14}$ s$^{-1}$ (Section \ref{sec:rotation}).
Figure \ref{fig:phys_c3} shows the physical structure at selected times in the model with the lower rotation rate.
Compared to the fiducial model, the envelope materials close to the cavity walls are denser at a given time.
As a result, stellar UV photons are shielded more efficiently and the UV radiation field is weaker in the envelope.
It is also seen that the outer disk radius is smaller than that in the fiducial model.
Figure \ref{fig:chem_c3} shows the distributions of H$_2$O ice abundance (left) and the $\hdo$ ice ratio (right) at $t = 0.3t_{\rm acc}$-0.9$t_{\rm acc}$ 
on a 1000 au scale, while Figure \ref{fig:chem_c3_zoom} shows the zoom-in views on disk scales.
Figure \ref{fig:uv_ab_case3} shows water ice abundances and the abundance ratios as functions of the normalized UV fluence defined by Equation (\ref{eq:ft}).
 Compared with the fiducial model, the UV fluence is significantly lower for the majority of the fluid parcels, and thus the water ice abundances tend to remain uniform.

\section{The ratio of $\ddo$ to $\hdo$ in a Class II turbulent disk chemical model}
\label{appendix:mixing}
\begin{figure*}
\resizebox{12cm}{!}{\includegraphics{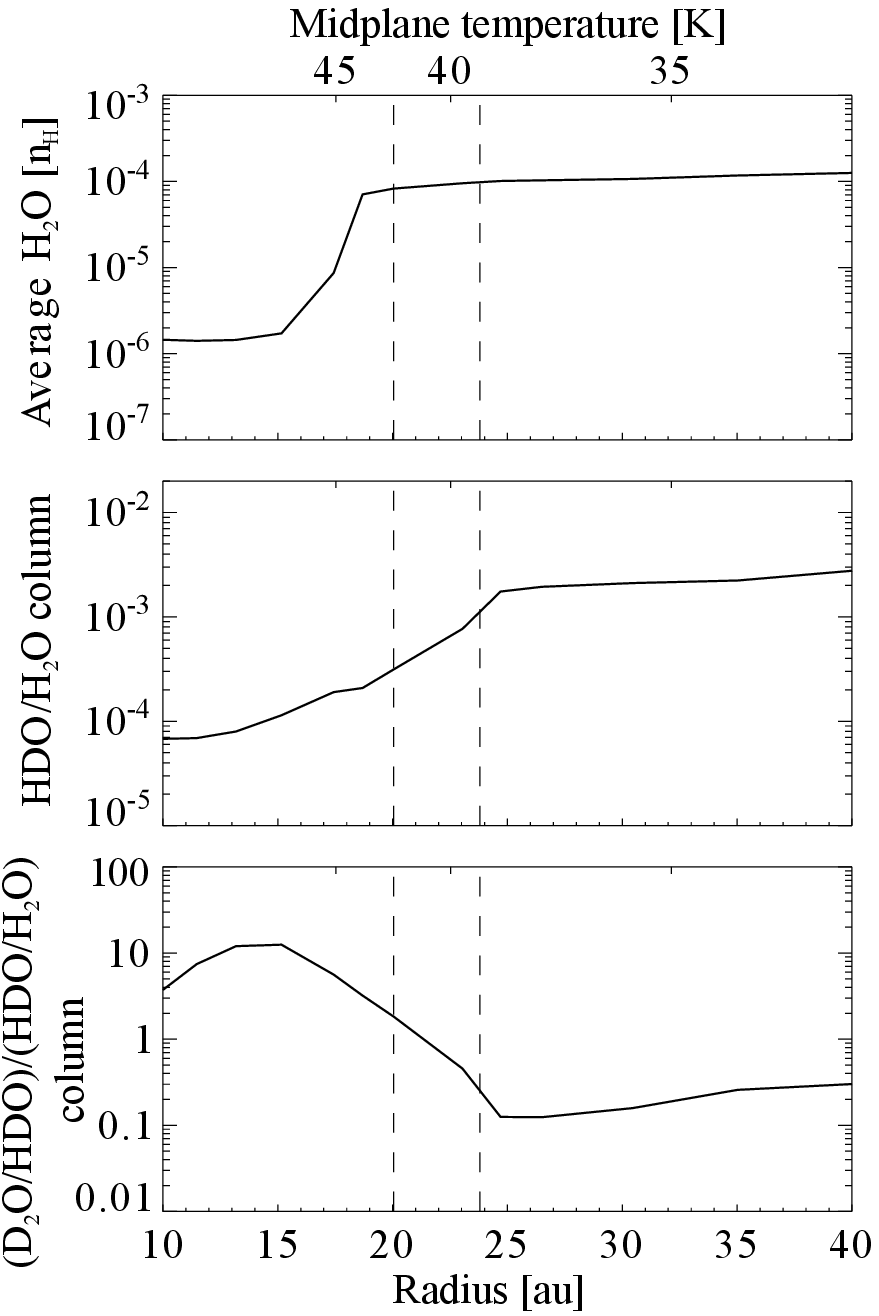}}
\caption{
Vertically averaged H$_2$O ice abundance (top), $\hdo$ ice column density ratio (middle), and the ice column density ratio of $\ddo$ to $\hdo$ (bottom) 
in a disk model around a typical T Tauri star after 1 Myr evolution.
The data are taken from \citet{furuya13} and \citet{furuya14}.
In regions between the dashed vertical lines, H$_2$O ice is abundant ($\sim$10$^{-4}$) and the cometary $\hdo$ ice ratio is reproduced.
The top label represents the temperature in the disk midplane.
}
\label{fig:mixing}
\end{figure*}

Figure \ref{fig:mixing} shows the vertically averaged H$_2$O ice abundance (top), the $\hdo$ ice column density ratio (middle), 
and the ice column density ratio of $\ddo$ to $\hdo$ (the bottom) in a disk model around a typical T Tauri star after 1 Myr chemical evolution.
The data are taken from an updated version of \citet{furuya13} model, presented in \citet{furuya14}.
The alpha parameter for the vertical diffusion coefficient, which decides the efficiency of vertical mixing of materials, is set to 10$^{-2}$.
Note that deuterium and ice chemistry in the turbulent disk model are limited (e.g., neither spin state chemistry nor ice layered structure are considered) 
compared to that in the present work. 
At radii 20-24 au, H$_2$O ice is abundant ($\sim$10$^{-4}$ [$n_{\rm H}$]) and the cometary $\hdo$ ice ratios \citep[$(3-11)\times 10^{-4}$;][and references therein]{mumma11} are reproduced.
In this region, the ratio of $\ddo$ to $\hdo$ ranges from $\sim$2 to $\sim$0.1.
The significant enhancement of the ratio of $\ddo$ to $\hdo$ at radii $\lesssim 20$ au is caused by the production of D$_2$O in the gas phase followed by freeze-out,
rather than by the icy grain surface chemistry.
In such regions, the H$_2$O ice abundance is low, since the warm dust temperatures reduce the reformation rate of H$_2$O ice \citep{furuya13}.

\end{appendix}

\end{document}